\documentclass[final,authoryear,3p,times]{elsarticle}

\usepackage{amssymb}
\usepackage{amsthm}
\usepackage{amsmath}
\usepackage{algpseudocode}
\usepackage[boxed]{algorithm2e}
\usepackage{float}
\usepackage{color}
\usepackage{geometry}
\usepackage{pdflscape}
\usepackage{tabulary}
\usepackage{changepage}
\usepackage{adjustbox}
\usepackage{rotating}
\usepackage{tabularx,longtable,supertabular,booktabs,dcolumn,multicol,multirow}
\usepackage[title]{appendix}
\usepackage{threeparttable}

\usepackage{lineno}

\usepackage{hyperref}
\hypersetup{
	colorlinks   = true,    
	urlcolor     = blue,    
	linkcolor    = blue,    
	citecolor    = blue      
}

\begin{document}

\begin{frontmatter}

\title{A New Spatial Count Data Model with Time-varying Parameters} 


\author[2]{Prasad Buddhavarapu \fnref{label1}}
\ead{prasad.buddhavarapu@utexas.edu}
\fntext[label1]{Department of Civil Architectural and Environmental Engineering,The University of Texas at Austin}

\author[1,2]{Prateek Bansal\fnref{label2}}
\fntext[label2]{Department of Civil and Environmental Engineering, Imperial College London}


\author{Jorge A. Prozzi\fnref{label1}}
\ead{prozzi@utexas.edu}

\address[1]{Corresponding author: \textcolor{blue}{prateek.bansal@imperial.ac.uk}}
\address[2]{Equal contribution.}

\begin{abstract}
Recent crash frequency studies incorporate spatiotemporal correlations, but these studies have two key limitations -- i) none of these studies accounts for temporal variation in model parameters; and ii) Gibbs sampler suffers from convergence issues due to non-conjugacy. To address the first limitation, we propose a new count data model that identifies the underlying temporal patterns of the regression parameters while simultaneously allowing for time-varying spatial correlation. The model is also extended to incorporate heterogeneity in non-temporal parameters across spatial units. We tackle the second shortcoming by deriving a  Gibbs sampler that ensures conditionally conjugate posterior updates for all model parameters. To this end, we take the advantages of P{\'o}lya-Gamma data augmentation and forward filtering backward sampling (FFBS) algorithm. After validating the properties of the Gibbs sampler in a Monte Carlo study, the advantages of the proposed specification are demonstrated in an empirical application to uncover relationships between crash frequency spanning across nine years and pavement characteristics. Model parameters exhibit practically significant temporal patterns (i.e., temporal instability). For example, the safety benefits of better pavement ride quality are estimated to increase over time.
\end{abstract}


\end{frontmatter}

\newpage

\section{Introduction}
\subsection{Background}
Traffic crashes are one of the main sources of fatalities in the United States. The National Highway Traffic Safety Administration reported 37,133 crash-related fatalities in the year 2017, which resulted in the economic cost of \$242 billion \citep{nhtsa2017}. These startling statistics call for new safety countermeasures and policies. To this end, spatial count data models have been adopted to uncover complex relationships between crash counts and influencing factors such as road conditions and geometric features. 

There are three main sources of \textit{unobserved heterogeneity} in crash frequency modeling and neglecting it may result in biased parameter estimates and an inaccurate policy guidance \citep{mannering2016unobserved}. First, the crash information collected from policy reports and other databases lacks several factors such as human behaviour, vehicle characteristics, and environmental conditions that can influence the likelihood of an accident. Such unobserved factors may introduce \textit{observation-specific variation} into the relationship between observed explanatory variables and crash count outcomes. Second, since accidents are rare events, they are generally aggregated over time (e.g. day, month, or year) and space (e.g., county or census tract) to ensure that each observation unit has adequate crash frequencies for statistical analysis \citep{lord2010statistical,mannering2014analytic, mannering2018temporal}. Previous studies have shown that crash counts in spatial units may be correlated due to resemblance in land use, weather, traffic laws, and driving behaviour \citep{liu2017exploring, li2019hierarchical}. Third, the parameters estimates can also exhibit temporal instability (or correlation) due to temporal changes in driver's decision-making, risk-taking behaviour, and cognitive biases \citep{mannering2018temporal}. These \textit{temporal correlations} between parameters of different time units are  another source of unobserved heterogeneity. 

To specify \textit{observation-specific variations} in the effect of the observed variables, model parameters are assumed to be random variables with various parametric and semi-parametric distributions. Dirichlet process mixture \citep{heydari2017using} and its parametric counterpart, a finite mixture of Gaussian distributions \citep{buddhavarapu2016modeling}, are the state-of-the-art mixing distributions. Both are discrete-continuous representations of heterogeneity where each observation has a probabilistic association with latent classes and a normal distribution is specified within each class. 

To specify the unobserved spatial dependence between observations, various specifications of spatial correlation have been explored in the literature -- intrinsic conditional autoregressive (ICAR) \citep{macnab2004bayesian, aguero2008analysis, wang2013poisson}, spatial autoregressive and spatial error model \citep{quddus2008modelling} , and geographic weighted Poisson regression \citep{hadayeghi2010development}. There is no consensus among researchers in terms of superiority of one specification over others and the choice of the specification is generally driven by computational convenience. In contrast to the abundant literature on modeling spatial correlation, only a handful of crash frequency studies account for spatiotemporal correlations. \cite{miaou2003roadway} first introduced spatiotemporal correlations to traffic crash frequency modeling by adopting a hierarchical Bayesian framework. In the frequentist setting, seminal work by \cite{castro2012latent} facilitates the incorporation of spatiotemporal correlation and observation-specific heterogeneity in count data models by recasting them as a restricted version of a generalized ordered response model. Some recent studies propose variants of spatiotemporal count data models, but most of them resort to the Bayesian estimation \citep{aguero2006spatial,truong2016spatiotemporal,
dong2016macroscopic,  liu2017exploring, ma2017multivariate,
cheng2018bayesian, liu2018using, li2019hierarchical}. This is perhaps because Markov Chain Monte Carlo (MCMC) methods are easier to implement in a canned software like OpenBUGS \citep{lunn2009bugs} or WinBUGS \citep{lunn2000winbugs}. 

\subsection{Research Gaps}
We identify two main limitations of the existing count data models with spatiotemporal correlations: 
\begin{itemize}
     \item \textit{Modeling:} none of these models incorporates temporal variation in model parameters, rather temporal correlation is specified in the link function after conditioning on observed covariates. Such specifications cannot model the temporal variation in parameters. In the absence on any coherent model, the temporal instability of parameters is often quantified by estimating crash count/severity data models for each period, followed by the hypothesis testing to evaluate whether parameters of consecutive periods are statistically different or not \citep{islam2020temporal,islam2020unobserved}. Such methods fail to account for the inherent dependence between parameters of consecutive time periods.          
    \item \textit{Estimation:} the studies relying on MCMC-based estimation use conventional Gibbs samplers, which do not have closed-form conditional marginal posteriors. Therefore, they have to embed the Metropolis-Hastings (MH) routine into a Gibbs sampler for posterior inference. This approach is prone to computational and convergence issues \citep{liu2017exploring}, and is highly sensitive to initial values \cite{liu2018using}. This is because the step size tuning in MH is challenging -- a small step size leads to high serial correlation and a large step size may not fully explore the posterior domain \citep{rossi2012bayesian}.
\end{itemize}

\subsection{Contributions}
To address these research gaps, we advance the specification of temporal correlation in spatial Negative-Binomial (NB) models and propose an efficient posterior inference routine. The proposed specification allows for temporal variation in NB parameters using dynamic linear models (DLMs) and temporal variation in spatial correlations by leveraging ICAR priors. DLMs provide a flexible structure, which not only accounts for cross-temporal correlations across regression coefficients, but also enables temporal variation in the coefficients of the auto-regressive process. The conventional Gibbs sampler for this model also suffers from the unavailability of non-conjugate priors for the NB likelihood. To this end, we add P{\'o}lya-Gamma-distributed auxiliary variables in the hierarchical structure of the specification to transform the NB likelihood into Gaussian distribution and the resulting conjugate structure provides a Gibbs sampler with closed-form posterior updates  \citep{polson_bayesian_2013}. The proposed P{\'o}lya-Gamma augmented Gibbs sampler circumvents the need for MH steps in MCMC simulation, which enables computationally-efficient and robust estimation. We first validate the inference procedure in a Monte Carlo study and illustrate its application in estimating crash counts of the contiguous road segments in the Houston area from 2003 to 2011. 

We also extend the proposed specification to additionally account for observation-specific heterogeneity in non-temporal parameters, where the heterogeneity is specified using a finite mixture of Gaussian distributions    \citep{buddhavarapu2016modeling}. A P{\'o}lya-Gamma-augmented Gibbs sampler is also derived for this extension.  

The remaining of this paper is organized as follows. The proposed specification and the Bayesian inference algorithm are described in sections \ref{sec:dev} and \ref{sec:inference}, respectively. Section \ref{sec:ext} outlines the changes in the original model specification due to inclusion of the observation-specific heterogeneity in non-temporal parameters and discusses corresponding modifications in the Gibbs sampler. Subsequently, a Monte Carlo study is presented in section \ref{sec:monte}, followed by the empirical study in section \ref{sec:empirical}.  Lastly, section \ref{sec:conc} concludes with the key findings, and highlights potential avenues for future research.

\section{Model Development} \label{sec:dev}
We analyze a crash count data across $T$ years from $n$ contiguous road segments. Let $y_{it}$ represents the crash count on $i^{th}$ road segment during $t^{th}$ year. Crash counts $\{{y_{it}: i \in \{1,2.....,n\}, t \in \{1,2.....,T\}}\}$ are assumed to be generated by a NB process with parameters $p_{it}$ and $r$. The site-specific attributes may be divided into two groups based on their effect on the respective crash count: time-invariant fixed parameters and time-varying parameters. Let $X^{F}_{it} = [x^{F}_{it1},x^{F}_{it2},....,z^{F}_{itg}]$ denote the $1 \times g$ attribute vector with fixed coefficients $\gamma = [\gamma_1,\gamma_2,...\gamma_g]$. Also, let $X^{D}_{it} = [X^{D}_{it1},X^{D}_{it2},.....,X^{D}_{itq}]$ denote a $1 \times q$ vector of attributes with dynamic parameters and $\theta_t = \{ \theta_{t1}, \theta_{t2},......,\theta_{tq}\}$ denote the vector of time-varying regression coefficients. Note that the matrices $X^F$ and $X^D$ are of $nT \times g$ and $nT \times q$ dimension, respectively. The proposed crash count model is described below:

\begin{equation} \label{eq:NBLH}
\begin{aligned}
y_{it} &\sim \text{NB}(r, p_{it}); \quad i \in \{1,2,...n\};  t \in \{1,2,...T\} \\
p_{it} &= \frac{\exp(\psi_{it})}{1+\exp(\psi_{it})}; \quad \psi_{it}= X^{F}_{it}\gamma + X^{D}_{it}\theta_{t} + \phi_{i}^{t}
\end{aligned}
\end{equation}

We consider all attributes to be time-varying, but time-invariant attributes can also have a time-varying effect and can be included in the current specification by repeating them across periods. The temporal variation of the regression coefficients $\theta_t$ is modeled as a dynamic linear model. The crash counts of contiguous road segments are likely to be spatially correlated and the magnitude of spatial correlation may change over time. The proposed specification allows for time-varying spatial correlation through time-specific spatial random effects. A vector of spatial random effects $\phi^{t}$ generated using ICAR prior structure is utilized to induce spatial correlations across the crash counts at time $t$. In subsequent subsections, we discuss the specification of time-varying parameters and ICAR prior structure, followed by summarizing the generative process of the proposed model.   

\subsection{Dynamic Linear Models} \label{sec:DLM}

Dynamic regression facilitates the variation of the parameters according to a specified state-space structure. For instance, dynamic linear models (DLMs) are formulated by assuming linear operators while specifying the system of equations. DLMs are extensively used in time series applications for extracting the underlying states that might be driving temporal changes in the outcome of interest. We assume that the following DLM structure generates the observed crash count time series.
\[
\zeta_t = F_t\theta_t + \nu_t; \quad \nu_t \sim \text{Normal}(0,V_t)
\]
\[
\theta_t  = G_t \theta_{t-1} + u_t;  \quad  u_t \sim \text{Normal}(0,W_t)
\] 

while inferring vector $\theta_{1:T}$ conditional on other model parameters, Bayesian implementation allows to pretend $\zeta_t$, instead of $y_t$ ($n \times 1$ vector of crash counts at time $t$), as the observed outcome at time $t$. A data augmentation technique is employed to transform $y_t$ into a multivariate Gaussian distributed random variable $\zeta_t$ (construction of $\zeta_t$ from $y_{t}$ is further discussed in subsection \ref{sec:FFBS}). The attribute matrix $F_t$ is generally constructed using the time-varying attributes (i.e., $X^{D}_t$, $n \times q$), but time-invariant attributes can be included by repeating them across time periods. The vector of observations $\zeta_t$ are generated by the latent parameter vector $\theta_t$ after transformation using the attribute matrix $F_t$ and adding a zero-centered multivariate Gaussian noise term $\nu_t$ with the covariance matrix $V_t$.

The latent parameter vector $\theta_t$ is assumed to be generated according to a linear state equation. $\theta_t$ is generated by the transformation of $\theta_{t-1}$ using the system operator matrix $G_t$ ($q\times q$) and adding a zero-centered multivariate Gaussian noise term $u_t$ with the covariance matrix $W_t$. $G_t$ may be designed such that $\theta_t$ is generated through an auto-regressive (AR-1) process. The DLM structure allows for specifying any other general auto-regressive structure. For instance, $G_t$ matrix may be specified such that a current state $\theta_{tk}$ is dependent on another previous state $\theta_{tk'}$ (where $k\neq k'$), rather than just on the last time period. In addition, $G_t$ may be designed as a time-varying system operator matrix; however, we assume it time-invariant. The proposed DLM framework is thus adequately flexible to investigate several underlying temporal patterns beyond the specific structure considered in this study.

\subsection{Time-varying intrinsic conditional autoregressive priors} \label{sec:car}
As mentioned earlier, we transform the NB likelihood into Gaussian likelihood by adding P{\'o}lya-Gamma-distributed auxiliary variables (more details in subsection \ref{sec:aug}). To exploit this conjugacy attained with the data augmentation, we assume Gaussian ICAR prior on spatial random effects. ICAR prior generates spatially correlated random effects based on a neighborhood or distance based correlation matrix. We utilize a neighborhood weight matrix $W$ defined as follows: $w_{ij} = 1/k$, if $i$ and $j$ are k-order neighbors. We denote $w_{i+}$ as the sum of $i^{th}$ row of the weight matrix. The spatial dependence is assumed to be proportional to the closeness of the neighboring road segments. The spatial correlation may vary with time, which we model by allowing for temporal variation in the parameter $\tau_t$. ICAR prior specifies the distribution of spatial random effect of $i^{th}$ road segment at time $t$ (i.e. $\phi_i^{t}$), conditional on spatial random effects of other road segments (i.e. $\phi_{-i}^{t}$): 

\begin{equation}
\phi_i^{t}|\phi_{-i}^{t} \sim N\left(\sum\limits_{j}^{}{\frac{w_{ij}}{w_{i+}}\phi_j^{t},\frac{\tau_t^2}{w_{i+}}}\right) 
\end{equation}

It is worth noting that the $\tau_t$ parameter does not quantify the strength of spatial correlation at time $t$ \citep[see section 3.3 of][for a detailed discussion]{banerjee_hierarchical_2004}. Total variation in mean crash count may be decomposed into unstructured heterogeneity and structured spatial variation; we use the proportion of variation due to spatial clustering as an estimate of strength of spatial correlation: $\alpha_{t} = \frac{\sigma_{\phi^{t}}}{\sigma_{\phi^{t}} + \sigma_{\epsilon^{t}}}$, where  $\sigma_{\phi^{t}}$ is the empirical standard deviation of posterior draws of spatial random effects, and $\sigma_{\epsilon^{t}}$ is the standard deviation of the unstructured random effects generated by the Gamma mixing in the NB model at time $t$. We construct the empirical posterior distribution of the $\alpha_{t}$ by computing $\sigma_{\phi^{t}}$ and $\sigma_{\epsilon^{t}}$ in each MCMC iteration to estimate the strength of spatial correlation.

\subsection{Generative Process} \label{sec:final}
To facilitate the Bayesian estimation, non-informative conjugate priors are imposed on the model parameters. The generative process of the proposed NB spatial model with time-varying parameters is summarised below. 

\[
y_{it} \sim \text{NB}(r, p_{it}); \quad i \in \{1,2,...n\};  t \in \{1,2,...T\}
\]

\[
p_{it} = \frac{1}{1+\exp(-\psi_{it})}; \quad \psi_{it}= X^{F}_{it}\gamma + X^{D}_{it}\theta_t + \phi_i^{t}
\]

\[
\theta_t  = G_t \theta_{t-1} + u_t; \quad u_t \sim \text{Normal}(0,W_{t})
\]
Where, 
\[
\theta_{t} = 
\begin{bmatrix}
       \theta_{1t} \\[0.3em]
       \theta_{2t} \\[0.3em]
       ...\\
			 \theta_{qt} 			
     \end{bmatrix} \quad 
G_{t} = 
\begin{bmatrix}
       \rho & 0 & .. &..& 0           \\[0.3em]
       0 & \rho &..&..& 0 \\[0.3em]
       .. & ..&..&..&.. \\
				0 & 0 & ..&..&\rho
     \end{bmatrix}	  \quad 
W_{t} = 
\begin{bmatrix}
       \sigma_{\theta 1}^2 & ..&..&..& 0 \\[0.3em]
       0 & \sigma_{\theta 2}^2 & ..&..& 0 \\[0.3em]
       ..& ..&..&..&..&\\
				0 & 0 &..&..& \sigma_{\theta q}^2
     \end{bmatrix}	
\]
\[
\gamma \sim \text{Normal}(s_0,S_0);  \quad  \theta_0 \sim \text{Normal}(m_0,C_0);  \quad \left\{\frac{1}{\sigma_{\theta k}^2}\right\}_{k=1}^{q} \sim \text{Gamma}(a_{\sigma}, b_{\sigma})
\]
\[
\phi_i^{t}|\phi_{-i}^{t} \sim N\left(\sum\limits_{j}^{}{\frac{w_{ij}}{w_{i+}}\phi_j^{t},\frac{\tau_t^2}{w_{i+}}}\right) 
\]
\[
\tau_t^{-2} \sim \text{Gamma}(c_0,d_0); \quad  r \sim \text{Gamma}(r_0, h);  \quad  h \sim \text{Gamma}(he_0,hf_0); 
\]

where Gamma$(\varphi,\chi)$ is Gamma distribution with mean $\frac{\varphi}{\chi}$. $\{ r, h, \gamma, \omega_{it}, \theta_{t},\sigma_{\theta k}^2, \phi_{i}^{t}, \tau_{t}\}_{\{\forall i,t, k\}}$ is a set of model parameters and $\{r_0, e_{0}, f_{0}, s_0, S_0,  m_0, C_0, c_0, d_0, a_{\sigma}, b_{\sigma} \}$ is a set of hyper-parameters. We set $\rho=1$ in this analysis. 

\section{Bayesian Inference} \label{sec:inference}
The model parameters are estimated using Markov Chain Monte Carlo (MCMC) simulation. As discussed earlier, analytical full conditional distributions are not available for NB regression models because their likelihood does not have a conjugate prior specification. In this section, we first discuss the intuition behind the P{\'o}lya-Gamma data-augmentation, which we use to address the non-conjugacy of NB regression. Subsequently, we discuss Forward Filtering Backward Sampling (FFBS) algorithm \citep{prado_time_2010}, which we employ for posterior inference of time-varying parameters. We also discuss how the P{\'o}lya-Gamma data augmentation facilitates the integration of the FFBS algorithm into the Gibbs sampler of the proposed model. The steps of the Gibbs sampler are summarized in algorithm \ref{algo:1}. 

\subsection{Data Augmentation} \label{sec:aug}
Data augmentation involves introducing latent random variables into the specification that are useful to derive analytically tractable full conditional posteriors. P{\'o}lya-Gamma data augmentation relies on adding P{\'o}lya-Gamma-distributed auxiliary variables. Conditional on these additional variables, the logistic likelihood translates into Gaussian likelihood. The same strategy works for the NB regression because it involves logistic likelihood. The details on the transformation of NB likelihood into Gaussian likelihood are provided in Appendix \ref{app:polya}. The conjugacy of the conditionally Gaussian likelihood and multivariate normal priors leads to analytically tractable full conditional posterior distributions for the non-temporal fixed parameters (i.e., $\gamma$). 

The P{\'o}lya-Gamma data augmentation also allows to construct a tractable analytical full conditional posterior for spatial random effects. Time-dependent ICAR priors are improper probability distributions on the vector of spatial random effects \citep{banerjee_hierarchical_2004}. However, the full conditional posterior turns out to be a proper probability distribution. We impose a sum-to-zero constraint for each time $t$ by recentering the draws of $\phi^{t}$ in each MCMC iteration; i.e. $\phi_i^{t} = \phi_i^{t} - \sum\limits_{i=1}^{n}{\phi_i^{t}}$ $\forall i \in \{1,2,.....,n\}; t \in \{1,2,.....,T\}$. 

We also employ another data augmentation technique proposed by \citet{zhou_lognormal_2012} to construct the full conditional distribution for the dispersion parameter $r$ (see Appendix B.1 for details). We consider an alternate representation of NB regression -- compound Poisson factorization, which introduces a Poisson-distributed latent indicator variable $L_{it}$ for each spatial unit $i$ and period $t$ in the model structure. The full conditional posterior of the dispersion parameter $r$ turns out to be Gamma distribution. 

\begin{algorithm}[!ht]
\SetAlgoLined
\SetKwFor{For}{for}{sample}{end for}%
\textbf{Initialization:}\\
Initialize parameters: $\{ r, h, \gamma,  \omega_{it}, \theta_{t}, \sigma_{\theta k}^2, \phi_{i}^{t}, \tau_{t} \}, \quad \forall i \in \{1,\dots,n\}, \forall t \in \{1,\dots,T\}, \forall k \in \{1,\dots,q\}$  \;
Set hyper-parameters: $\{r_0, e_{0}, f_{0}, s_0, S_0, m_0, C_0, c_0, d_0, a_{\sigma}, b_{\sigma} \}$    \;
\For{1 \KwTo max-iteration}{
i) $ r \lvert \text{\textendash} \sim  \text{Gamma}\left(r_0+\sum\limits_{t=1}^{T}\sum\limits_{i=1}^{N}L_{it}, h-\sum\limits_{t=1}^{T}\sum\limits_{i=1}^{N}\ln(1-p_{it})\right)$  (see equation \ref{eq:L} for the details of $L_{it}$)\;
ii) $h \lvert \text{\textendash} \sim  \text{Gamma}(r_{0} + he_{0},  r + hf_{0})$ \;
iii) $\gamma \lvert \text{\textendash} \sim \text{Normal}\left(V_{\gamma}\left(\sum\limits_{i=1}^{n} {X^F_i}^{\prime} \Omega_i^{-1} (z_i-\phi_i  - X_i^D \theta_{t}) + S_0^{-1}s_0\right), V_{\gamma} \right)$, where $ V_{\gamma} =   \left(\sum\limits_{i=1}^{n} {X^F_i}^{\prime} \Omega_i^{-1} X^F_i + S_0^{-1}\right)^{-1}$ \;
iv) $\left\{\omega_{it} \lvert \text{\textendash}\right\}_{\forall i,t} \sim \text{PG}(y_{it}+r, \psi_{it})$ \;
\vspace{0.3cm}
v) $\{\theta_{t}\}_{t=1}^{T}$ using forward filtering and backward smoothing (appendix B.5)\; 
\vspace{0.3cm}
Forward filtering: \\
 \For{$t$ in $1:T$}{
    Posterior at $t-1$: $\theta_{t-1} \lvert  D_{t-1} \sim \text{Normal}(m_{t-1},C_{t-1})$ \;
    Prior at $t$: $\theta_{t} \lvert  D_{t-1} \sim \text{Normal}(G_t m_{t-1},G_t C_{t-1} G_t^{\prime} + W_t)$ \;
    Predictive at $t$: $\zeta_t \lvert  D_{t-1} \sim \text{Normal}(F_t a_t, F_t R_t F_t^{\prime} + \Omega_t)$ \;
    Posterior at $t$: $\theta_{t} \lvert D_{t}  \sim \text{Normal}(a_t + R_t F_t^{\prime} Q_t ^{-1}(\zeta_t-f_t), R_t - R_t F_t^{\prime} Q_t^{-1} F_t R_t)$ ;
    \\}
\vspace{0.3cm}
Backward smoothing: \\
\For{$t$ in $(T-1):1$}{
$\theta_{t} \lvert \theta_{t+1},D_{t} \sim \text{Normal}( m_t + C_t G_{t+1}^{\prime} R_{t+1}^{-1} (h_{t+1} - a_{t+1}), C_t - C_t G_{t+1}^{\prime} R_{t+1}^{-1} (R_{t+1} - B_{t+1}) R_{t+1}^{-1} G_{t+1}^{\prime} C_t);$ \\ 
}

vi) $\left\{\frac{1}{\sigma_{\theta k}^2}\right\}_{k=1}^{q} \lvert \text{\textendash}  \sim \text{Gamma} \left( a_{\sigma} +\frac{1}{2} (T-1), \frac{1}{b_{\sigma}+ \frac{1}{2}\sum\limits_{t=2}^{T} {(\theta_{t}-G_t[k,k] \theta_{t-1})}^2}\right)$\;
vii) $\left\{\phi_i^{t} \lvert \text{\textendash}\right\}_{\{\forall i,t\}} \sim \text{Normal}\left( \left(\omega_{it}+\frac{w_{i+}}{\tau_t^2}\right)^{-1} \left([z_{it}-(X^{F}_{it}\gamma  + X^{D}_{it}\theta_{t})]\omega_{it}+\left(\sum\limits_{j} w_{ij}\phi_j^{t}\right)\frac{1}{\tau_t^2}\right), \left(\omega_{it}+\frac{w_{i+}}{\tau_t^2}\right)^{-1} \right)$ \; 
viii) $\left\{\tau_t^{-2} \lvert \text{\textendash} \right\}_{t=1}^{T} \sim \text{Gamma}\left(c_0+\frac{n}{2}, d_0+\sum\limits_{i=1}^{n}\frac{w_{i+}}{2}\left[ \phi_{i}^{t} - \sum\limits_{j} \frac{w_{ij}}{w_{i+}} \phi_{j}^{t} \right]^{2}\right)$ \;
ix) Compute spatial correlation: $\{\alpha_{t}\}_{t=1}^{T} = \frac{\sigma_{\phi^{t}}}{\sigma_{\phi^{t}} + \sigma_{\epsilon^{t}}}$.
}
 \caption{Gibbs sampler for the spatial negative binomial model with time-varying parameters.}
 \label{algo:1}
 \end{algorithm}

\subsection{Forward Filtering Backward Sampling (FFBS)} \label{sec:FFBS}

We adopt FFBS algorithm (originally proposed by \citet{fruhwirth-schnatter_data_1994,carter_gibbs_1994}) for posterior sampling of time-varying regression parameters $\theta_{1:T}$. FFBS algorithm simultaneously produces posterior draws of the state vector $\theta_{1:T}$ through forward sampling followed by backward smoothing in each MCMC iteration. FFBS algorithm is not capable of handling a non-linear model (such as NB regression) \citep{windle_polya-gamma_2013}. However, the transformation of the NB-distributed crash counts $y_{it}$ into a conditionally Gaussian distributed data vector $z_{it} = \frac{y_{it}-r}{2\omega_{it}}$ using P{\'o}lya-Gamma data augmentation facilitates the adoption of FFBS in the proposed non-linear model. 

The NB likelihood shown in Equation \ref{eq:NBLH} can be equivalently written as (see Appendix \ref{app:polya} for details): 
\[
z_{it} \lvert \omega_{it}, \text{\textendash} \sim \text{Normal}(\psi_{it},\omega_{it}^{-1})
\]
Where, $\psi_{it}= X^{F}_{it}\gamma + X^{D}_{it}\theta_{t} + \phi_{i}^{t}$ and $\omega_{it}$ is a P{\'o}lya-Gamma distributed auxiliary variable. Now, the evolution equations of the system may be written as follows using the transformed data $z_t$. 
\[
z_t = X^{F}_{t}\gamma +  X^{D}_{t}\theta_{t} + \phi^{t} + \nu_t
\]
\[
z_t - X^{F}_{t}\gamma  - \phi^{t} = X^{D}_{t}\theta_{t}  + \nu_t
\]
\begin{equation}\label{eq:zeta_t}
\zeta_t = F_t \theta_t + \nu_t; \quad \nu_t \sim \text{Normal}(0,\Omega_t)
\end{equation}
\[
\Omega_t = 
\begin{bmatrix}
       \frac{1}{\omega_{1t}} & ....& 0 & 0 \\[0.3em]
       0 & \frac{1}{\omega_{2t}} & .... & 0  \\[0.3em]
       0 & 0 & .... & \frac{1}{\omega_{nt}}
     \end{bmatrix}	
\]
\[
\theta_t = G_t \theta_{t-1} + u_t;  \quad  u_t \sim \text{Normal}(0,W_t)
\]

where $\zeta_t =z_t - X^{F}_{t}\gamma -  \phi^{t}$ and $ F_t = X^{D}_{t}$. The above augmented specification matches with a traditional dynamic linear model.

\section{Extension: Inclusion of Random Parameters} \label{sec:ext}
We first discuss the required modifications in the original model specification to include non-temporal random parameters in section \ref{sec:modsp}. Subsequently, we highlight key modifications in the Gibbs sampler of the extended model in section \ref{sec:modgs}. 

\subsection{The Modified Specification}\label{sec:modsp}
Let $X^{R}_{it} = [X^{R}_{it1},X^{R}_{it2},....,X^{R}_{ith}]$ denote the $1 \times h$ attribute vector with time-invariant random parameters $\beta_i = [\beta_{i1},\beta_{i2},...\beta_{ih}]$. After introducing non-temporal random parameters, only link function $\psi_{it}$ in the original model (see Equation \ref{eq:NBLH}) is modified as: 

\begin{equation}
    \psi_{it}= X^{F}_{it}\gamma + X^{R}_{it}\beta_{i} + X^{D}_{it}\theta_{t} + \phi_{i}^{t}
\end{equation}

Following \cite{buddhavarapu2016modeling}, we consider a finite mixture of multivariate normal distributions on time-invariant random parameter $\beta_i$. If $\mu_c$ and $\Sigma_c$ are mean vector and covariance matrix corresponding to $c^{th}$ component and $\eta_c$ is a weight of the $c^{th}$ component in the mixture of $C$ components, the flexible discrete-continuous distribution on $\beta_i$ is represented as follows:

\begin{equation}\label{eq:beta mixture}
\beta_i \sim \sum\limits_{c=1}^{C} \eta_c \text{Normal}(\mu_c,\Sigma_c); \quad \sum\limits_{c=1}^{C} \eta_c = 1,
\end{equation}

Due to change in specification of $\psi_{it}$, $\zeta_{t}$ in equation \ref{eq:zeta_t} of the augmented dynamic linear model also changes to $\zeta_t =z_t - X^{F}_{t}\gamma - X^{R}_{t}\beta_{i} - \phi^{t}$. Apart from these modifications, we add the following priors to the the generative process of the original model as presented in section \ref{sec:final}: 
\[
\eta_c \sim \text{Dirichlet}(\alpha_0,...\alpha_0); 
\]
\[
\mu_c \sim \text{Normal}(b_0,B_0); \quad \Sigma_c^{-1}  \sim \text{Wish}(\nu_0,V_0);  
\]

where $Wish(\xi,\Xi)$ is Wishart distribution with mean $\xi \Xi$.  Thus, $\{ r, h, \gamma,  \beta_{i}, \mu_{c}, \Sigma_{c}, \eta_{c}, \omega_{it}, \theta_{t}, \sigma_{\theta k}^2, \phi_{i}^{t}, \tau_{t} \}_{\{\forall i,t,c,k \}} $ is a set of model parameters and $\{r_0, e_{0}, f_{0}, s_0, S_0, b_0, B_0, \nu_0, V_0, \alpha_0, m_0, C_0, c_0, d_0, a_{\sigma}, b_{\sigma} \}$ is a set of hyper-parameters for the extended specifications. We set $\rho=1$ in this analysis.

\subsection{The Modified Gibbs Sampler}\label{sec:modgs}
Since the inclusion of non-temporal random parameters in the model specification adds just another layer of parameters in the hierarchy, the data augmentation techniques used for the original specification are applicable for the extended specification. In fact, as a consequence of P{\'o}lya-Gamma data augmentation, the parameters associated with the mixing distribution of the additional non-temporal random parameters (i.e., $\mu_{1:C}, \Sigma_{1:C}, \eta_{1:C}$) also attain conjugate posterior updates. 

Conditional posterior distributions of the original model parameters (except $\gamma$, the non-random parameter) either remain unaffected or are slightly modified after addition of non-temporal random parameters. We develop a blocked Gibbs sampler for the extended specification to improve the mixing of Markov chains. The detailed derivation of the conditional distributions of model parameters for the extended specification are provided in Appendix \ref{app:gibbs} and the steps of the Gibbs sampler are summarized in Algorithm \ref{algo:2}. Conditional distributions for original model parameters presented in Algorithm \ref{algo:1} can be easily retrieved from those derived for the extended specification due to nesting of the original specification within the extended specification.

\section{Monte Carlo Study}\label{sec:monte}

Before applying the proposed dynamic spatial NB (DSNB) model in an empirical application, validation of the finite sample and convergence properties of the Gibbs sampler is important. To this end, we simulate crash count data using the data generating process (DGP) of the DSNB specification and estimate the marginal posterior distributions of model parameters using the derived Gibbs sampler. 

\subsection{Data Generating Process}\label{sec:dgp}

We assume a highway network of 1000 road segments (i.e., $n=1000$) with crash count data across a span of 10 years (i.e. $T=10$). The link function has three covariates with time-invariant effect $\gamma$ and three covariates with time-varying effect $\theta_{t}$. Whereas features with time-invariant effects $X_{it}^{F}$ are generated from a multivariate normal distribution, features with time-varying effects $X_{it}^{D}$ are generated by assuming a series of correlated draws utilizing a sinusoidal temporal pattern with Gaussian random noise. A binary spatial weight matrix is created assuming that the crash counts of a given road segment are spatially correlated with those of 4 contiguous road segments on each side along the highway. The following true parameters are set prior to generating the intermediate model parameters: $\gamma$, $\sigma_{\theta\{1:q\}}^{2}$, $\tau_{1:T}$, $G_{1:T}$, and $r$. Subsequently, we generate $\{{\phi_i}^{t}\}_{i,t}$ and $\theta_{1:T}$ according to the spatial and dynamic linear models of the proposed DSNB specification. Lastly, $p_it$ is computed for each road segment $i$ at year $t$, followed by generating crash count $y_{it}$ using a negative binomial likelihood with parameters $p_it$ and $r$. We choose true model parameters in such a way that the distribution of the simulated crash counts approximately matches with that of the empirical crash data considered in this study. 

\subsection{Implementation, and Diagnostics} \label{sec:impl}

We write our own code to implement the Gibbs sampler (summarized in Algorithm \ref{algo:1}) on the simulated data in R software \citep{r_core_team_r:_2020}. We write several components of the code in Rcpp package \citep{eddelbuettel_rcpp:_2017} to gain additional computational advantages. To be specific, Rcpp implementation accelerates the computation of these components by a factor of 10 or more. We employ \emph{BayesLogit} package to sample efficiently from P{\'o}lya-Gamma-distributed random variables \citep{polson_bayeslogit_2012}. Around 2000 burn-in draws were deemed sufficient to attain stationary distribution, and the marginal posterior of model parameters were estimated based on subsequent 1000 MCMC draws. We carry out simulations on a Macintosh machine with Intel Core i5 CPU with 2.7 GHz and 8GB RAM. An average run time of around 6 hours is required to take 3000 MCMC draws from the joint distribution. 

To demonstrate the convergence of the Gibbs sampler, we report Geweke diagnostic statistic \citep{geweke1991evaluating}. The test takes two non-overlapping portions of the Markov chain, and performs a two-means Z-test to check for convergence. Moreover, we use the following metrics to assess the efficacy of the proposed Gibbs Sampler in recovering the true model parameters: 
\begin{itemize}
    \item \textit{Mean Absolute Bias (MAB)} =  $\left| \text{True parameter value - Posterior mean} \right|$
    \item \textit{Absolute Percentage Bias (APB)} = $\left| \frac{\text{True parameter value - Posterior mean} }{\text{True parameter value}} \right|\times 100$
    \item \textit{95\% credible interval coverage}: a binary indicator, which is 1 if the true parameter lies in the estimated 95\% credible interval. 
\end{itemize}

\subsection{Results of the Simulation Study}\label{sec:result_sim}

\begin{figure}[ht!]
\centering
\includegraphics[width=0.7\textwidth]{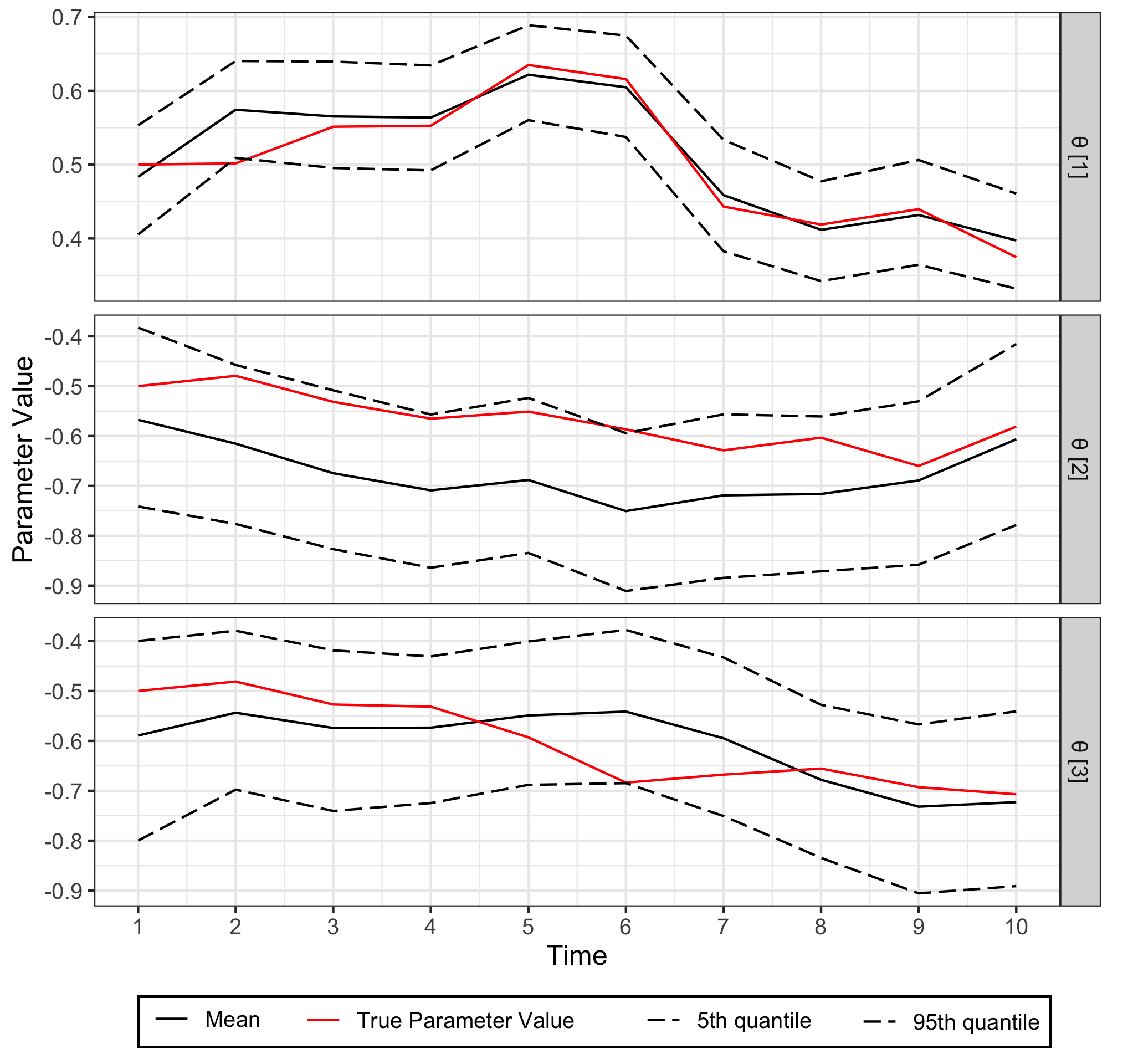}
\vspace{-0.4cm}\caption{Simulation results for dynamic spatial NB (DSNB) model: time-varying parameters}
\label{fig:sim_results_temporal}
\end{figure}

\begin{figure}[ht!]
\centering
\includegraphics[width=0.6\textwidth]{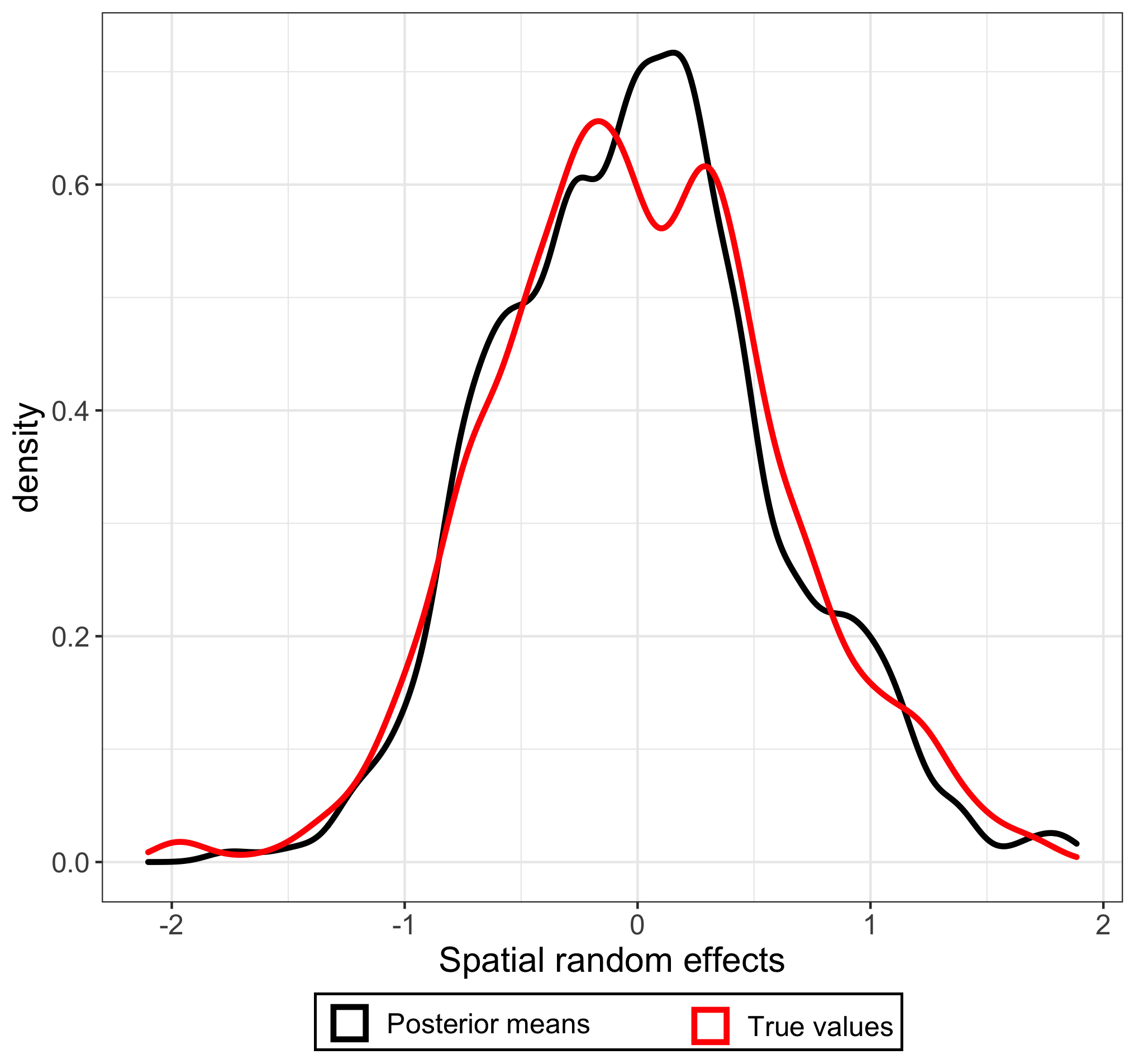}
\vspace{-0.4cm}\caption{Simulation results for dynamic spatial NB (DSNB) model: spatial random effects pooled across all years}
\label{fig:sim_results_spatial}
\end{figure}

Table \ref{tab:sim_results} summarizes the posterior statistics of model parameters, aforementioned performance metrics, and convergence diagnostics. The APB values of all model parameters range between 0.75\% and 28.39\%, and all key model parameters are captured within the estimated 95\% posterior credible intervals. To illustrate the recovery of time-varying parameters graphically, we plot the recovered posterior statistics of these parameters alongside their true values across ten years in Figure \ref{fig:sim_results_temporal}. We also superimpose the distribution of posterior means of road segment-specific spatial random effects on the true values used in generating crash counts in Figure \ref{fig:sim_results_spatial}. The resemblance in both plots ensures the efficacy of the proposed Gibbs sampler in recovering spatial random effects. 

\begin{figure}[ht!]
\centering
\includegraphics[width=0.6\textwidth]{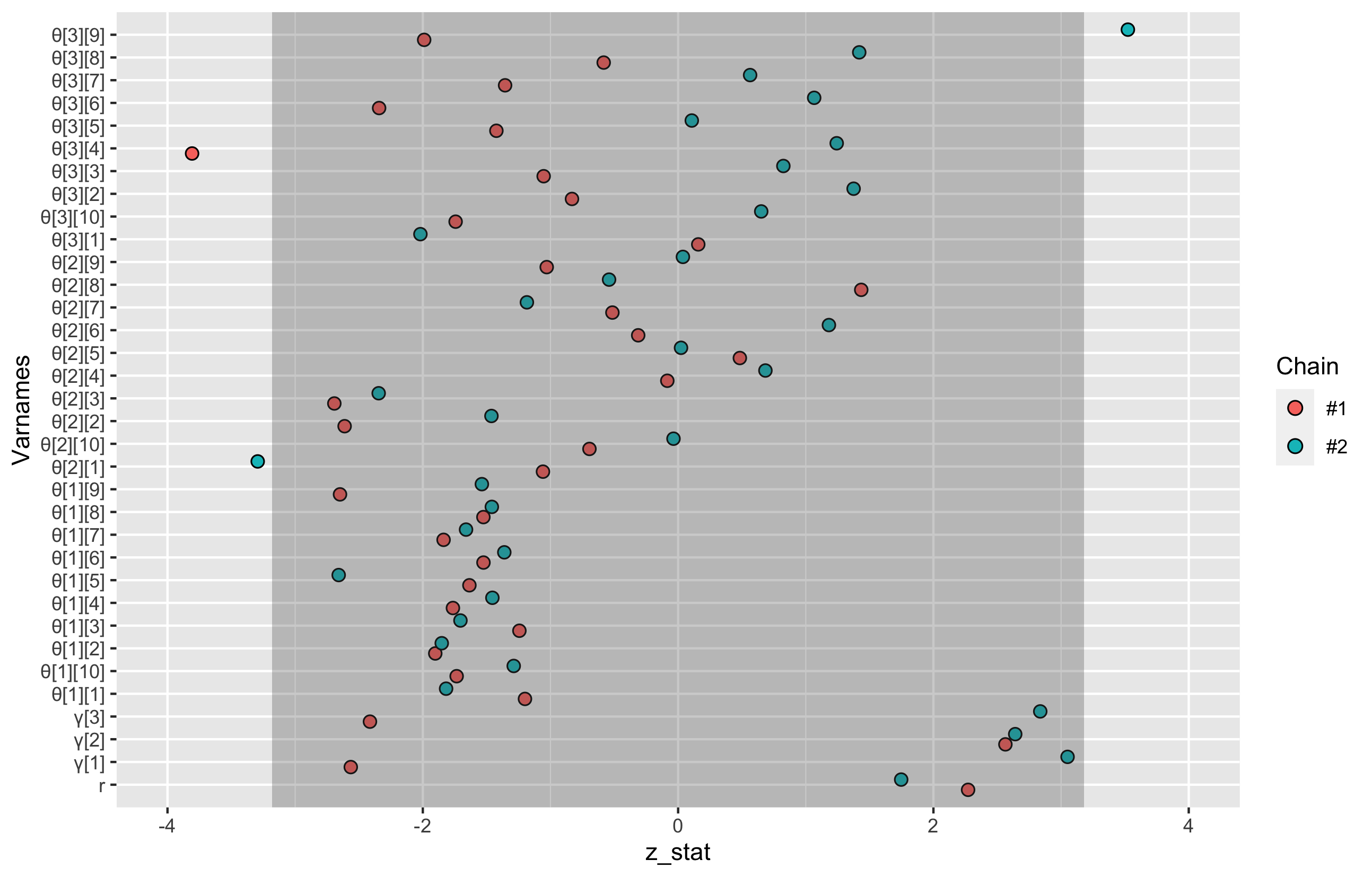}
\vspace{-0.4cm}\caption{Simulation results: Geweke diagnostic statistic}
\label{fig:geweke}
\end{figure}

We choose a 95\% threshold for hypothesis testing using Geweke statistic. In testing, we use Bonferroni correction \citep{napierala2012bonferroni} because we do multiple testing, i.e. individually check convergence of Markov chains for all 34 model parameters of interest (30 time varying, 3 time-invariant, and 1 dispersion parameter). The Geweke statistic of model parameters are shown in Figure \ref{fig:geweke} -- Z-score of the most of the model parameters is between -3.18 and 3.18 (95\% confidence interval thresholds with Bonferroni correction), suggesting that Markov chains have attained stationarity and the sampler has converged. 

In summary, convergence diagnostics and parameter recovery metrics of all the relevant parameters, i.e. the ones that are used as input to devise safety policies, indicate that the presented Gibbs sampler in Algorithm \ref{algo:1} is appropriate for posterior inference in the DSNB model and can be used in empirical studies.

\begin{table}[ht!]
  \centering
  \scriptsize
  \caption{Results of the simulation for dynamic spatial NB (DSNB) model}
    \resizebox{1\textwidth}{!}{\begin{tabular}{rccccccc}
    \toprule
    \multicolumn{1}{c}{\textbf{Parameter }} & \textbf{True value} & \textbf{Posterior mean} & \textbf{2.5\%-quantile} & \textbf{97.5\%-quantile} & \textbf{MAB} & \textbf{APB} & \textbf{95\%-coverage} \\ 
    \midrule
    \multicolumn{8}{c}{Time-invariant parameters} \\
    \midrule
	\multicolumn{1}{l}{$\gamma_1$} & 0.200 & 0.204 &  0.188  &  0.220  &  0.004  &    2.04\% & 1\\
    \multicolumn{1}{l}{$\gamma_2$} & 0.100 & 0.095 &  0.071  &  0.117  & 0.005  &   4.98\% & 1 \\
    \multicolumn{1}{l}{$\gamma_3$} & -0.100 & -0.089 & -0.103  & -0.073  &  0.011  &  11.40\% & 1\\
    \multicolumn{1}{l}{$r$} & 1.50 & 1.511 &  1.430  &  1.597  &  0.011  &    0.75\% & 1\\
    \midrule
    \multicolumn{8}{c}{Time-varying parameters} \\
    \midrule
    \multicolumn{1}{l}{$\theta_{1}[1]$} & 0.500 &    0.483 &   0.388  &     0.563 & 0.017 &  3.32\% & 1\\
    \multicolumn{1}{l}{$\theta_{2}[1]$} & 0.502 &    0.574 &   0.498  &     0.654 &   0.073 &  14.47\% & 1\\
    \multicolumn{1}{l}{$\theta_{3}[1]$} & 0.551 &    0.565 &   0.483  &     0.654 &   0.014 &   2.53\% & 1\\
    \multicolumn{1}{l}{$\theta_{4}[1]$} & 0.553 &    0.564 &   0.477  &     0.650 &   0.011 &   2.00\% & 1\\
    \multicolumn{1}{l}{$\theta_{5}[1]$} & 0.635 &    0.622 &   0.549  &     0.701 &  0.013 & 2.08\% & 1\\
    \multicolumn{1}{l}{$\theta_{6}[1]$} & 0.616 &    0.605 &   0.526  &     0.689 &  0.011 &  1.80\% & 1\\
    \multicolumn{1}{l}{$\theta_{7}[1]$} & 0.443 &    0.459 &   0.366  &     0.548 &   0.016 &   3.51\% & 1\\
    \multicolumn{1}{l}{$\theta_{8}[1]$} & 0.419 &    0.412 &   0.325  &     0.490 &  0.007 & 1.72\% & 1\\
    \multicolumn{1}{l}{$\theta_{9}[1]$} & 0.440 &    0.432 &   0.349  &     0.518 &  0.008 &  1.79\% & 1\\
    \multicolumn{1}{l}{$\theta_{10}[1]$} & 0.375 &    0.397 &   0.312  &     0.471 &   0.023 &   6.07\% & 1\\
    
    \multicolumn{1}{l}{$\theta_{1}[2]$} & -0.500 &   -0.567 &  -0.768  &    -0.349 &  0.067 &  13.50\% & 1\\
    \multicolumn{1}{l}{$\theta_{2}[2]$} & -0.479 &   -0.615 &  -0.796  &    -0.431 &  0.136 &  28.39\% & 1\\
    \multicolumn{1}{l}{$\theta_{3}[2]$} & -0.531 &   -0.675 &  -0.862  &    -0.473 &  0.143 &  26.93\% & 1\\
    \multicolumn{1}{l}{$\theta_{4}[2]$} & -0.565 &   -0.709 &  -0.896  &    -0.530 &  0.144 &  25.46\% & 1\\
    \multicolumn{1}{l}{$\theta_{5}[2]$} & -0.551 &   -0.688 &  -0.875  &    -0.477 &  0.137 &  24.89\% & 1\\
    \multicolumn{1}{l}{$\theta_{6}[2]$} & -0.587 &   -0.751 &  -0.946  &    -0.571 &  0.164 &  27.94\% & 1\\
    \multicolumn{1}{l}{$\theta_{7}[2]$} & -0.629 &   -0.719 &  -0.912  &    -0.532 &  0.090 &  14.35\% & 1\\
    \multicolumn{1}{l}{$\theta_{8}[2]$} & -0.603 &   -0.716 &  -0.918  &    -0.536 &  0.113 &  18.71\% & 1\\
    \multicolumn{1}{l}{$\theta_{9}[2]$} & -0.660 &   -0.689 &  -0.893  &    -0.497 &  0.029 &   4.41\% & 1\\
    \multicolumn{1}{l}{$\theta_{10}[2]$} & -0.581 &   -0.606 &  -0.822  &    -0.381 &  0.025 &   4.36\% & 1\\
    
    \multicolumn{1}{l}{$\theta_{1}[3]$} & -0.500 &   -0.589 &  -0.831  &    -0.361 &  0.089 &  17.83\% & 1\\
    \multicolumn{1}{l}{$\theta_{2}[3]$} & -0.481 &   -0.543 &  -0.717  &    -0.346 &  0.062 &  12.99\% & 1\\
    \multicolumn{1}{l}{$\theta_{3}[3]$} & -0.527 &   -0.574 &  -0.770  &    -0.387 &  0.047 &   8.92\% & 1\\
    \multicolumn{1}{l}{$\theta_{4}[3]$} & -0.531 &   -0.573 &  -0.753  &    -0.393 &  0.042 &   7.93\% & 1\\
    \multicolumn{1}{l}{$\theta_{5}[3]$} & -0.593 &   -0.549 &  -0.711  &    -0.370 &   0.044 &  7.40\% & 1\\
    \multicolumn{1}{l}{$\theta_{6}[3]$} & -0.684 &   -0.541 &  -0.708  &    -0.328 &   0.142 & 20.83\% & 1\\
    \multicolumn{1}{l}{$\theta_{7}[3]$} & -0.668 &   -0.595 &  -0.786  &    -0.395 &   0.073 & 10.91\% & 1\\
    \multicolumn{1}{l}{$\theta_{8}[3]$} & -0.655 &   -0.678 &  -0.865  &    -0.503 &  0.023 &   3.44\% & 1\\
    \multicolumn{1}{l}{$\theta_{9}[3]$} & -0.693 &   -0.732 &  -0.935  &    -0.540 &  0.039 &   5.64\% & 1\\
    \multicolumn{1}{l}{$\theta_{10}[3]$} & -0.707 &   -0.723 &  -0.932  &    -0.511 &  0.016 &   2.23\% & 1\\
    \midrule
    \multicolumn{3}{l}{Deviance Information Criterion (DIC)} & 39263.25 \\
    \multicolumn{3}{l}{Number of MCMC iterations} & 3000 \\
    \multicolumn{3}{l}{Number of burn-in iterations} & 2000 \\
    \multicolumn{3}{l}{Number of simulated road segments} & 1000 \\
    \multicolumn{3}{l}{Number of simulated years of data} & 10 \\
    \multicolumn{3}{l}{Geweke convergence diagnostics} & See Figure \ref{fig:geweke}\\
    \bottomrule
    \end{tabular}}%
  \label{tab:sim_results}%
\end{table}%

\clearpage

\section{Empirical Analysis} \label{sec:empirical}

This section presents an application of the proposed DSNB model in understanding the relationship between road attributes and crash counts using panel data. Quantifying these relationships help in designing new safety-countermeasures. Studying the evolution of the relationship of road attributes with crash outcomes enables the development of informed safety improvement strategies.

\subsection{Data description}

In this empirical analysis, we model historical crash counts of contiguous freeway road segments of a metropolitan road network from Houston city, USA, which has grade-separated freeway junctions. We use data from eleven different road facilities across a span of 9 years (2003 to 2011). We source crash counts from the publicly available motor vehicle Crash Record Information System (CRIS) database maintained by the local Department of Transportation (DOT). We geographically map individual crash occurrences to the respective road segments and subsequently aggregate them temporally to obtain the annual crash counts for each road segment. We further integrate the crash data with road condition management databases, which track several road-specific attributes of the road transportation network. This integration helps in linking the annual crash counts with the respective annually aggregated road-segment-specific attributes. More details about the data collection and processing are presented in \citep{buddhavarapu2015bayesian}.

The attributes of road segments include traffic volumes, various geometric features, pavement surface characteristics and structural distresses, and road locations. Table \ref{tab:desc_time} reports annual summary statistics of these time-varying and time-invariant attributes of road segments and the key trends are discussed below. An overall decreasing trend of mean crash counts during the study period indicates that the study network witnessed safety improvements over time. Annual average daily traffic (AADT) increased on an average, while the proportion of truck traffic remained constant for the initial few years, followed by a slight reduction during the last part of the study period. The average speed limit across the road segments slightly changed from 2003 to 2004 and remained consistent until the end of the study period. The average number of lanes and shoulder widths did not vary much during the study period, indicating no significant changes to the road network in terms of road widening. The ride quality of a road (measured by international roughness index, IRI) and road distress index indicate that the road condition of the study network (on an average) improved slightly during the analysis period. This improvement reflects the efficacy of the maintenance efforts of DOT in managing the road network. The proportion of continuously reinforced concrete pavements (CRCP) also exhibits an increasing trend. About 45\% of the road segments belong to interstate highways (IH), and the rest of the road segments belong to state highways (SH) and US highways. The reduction in the proportion of the road segments with rural area flag indicates urban sprawl in the study area.

\subsection{Model estimation}
We have established the statistical properties of the Gibbs sampler for the DSNB model in the simulation study in section \ref{sec:monte}. In this empirical study, we follow the same specification and procedure to implement the Gibbs sampler as used in the simulation study (see section \ref{sec:impl} for details). For the model selection, i.e. to identify a set of road attributes that explain the variation in the crash counts, we test various model specifications and select the one with the lowest Deviance Information Criterion (DIC) \citep{spiegelhalter_bayesian_2002}). 

To demonstrate the importance of accounting for temporal variation in parameters, we also estimate another constrained model specification where we consider all parameters to be time-invariant. In other words, in addition to a DSNB specification, we also estimate a corresponding spatial NB (SNB) model.\footnote{Whereas we have extended the proposed DSNB model to account for unobserved heterogeneity in time-invariant parameters in section \ref{sec:ext} and also derived its Gibbs sampler in Algorithm \ref{algo:2}, we do not adopt this specification in this application because such effects are hard to empirically identify along with time-varying parameters using the sample size of the empirical study. The likelihood of empirically identifying such heterogeneous effects would improve with the increase in the number of road segments and the repeated observations in the sample.}

Tables \ref{tab:dynpost} reports the posterior summaries of DSNB model parameters. Since presenting all temporal parameters of DSNB in tabular format results in a very long table, we only report variance of the AR(1) process for each of these parameters in Table \ref{tab:dynpost}. However, we plot the posterior summaries of time-varying parameters and spatial correlation (see Section \ref{sec:car} for discussion on spatial correlation) of DSNB model in Figures \ref{fig:emp_results} and \ref{fig:sp_cor}, respectively, while juxtaposing the posterior estimates of the SNB specification for comparison. 
To perform statistical inference on model parameters, we report mean, standard deviation, and 95\% credible interval limits of the estimated posterior distributions. The posterior mean highlights the magnitude of the model parameters, while credible intervals provide insights on proximity to any null hypothesis value of interest. In the next subsection, we provide a detailed discussion on the estimation results and their implications to safety management.

\begin{landscape}
\vspace*{\fill}
\begin{table}[H]
\scriptsize
\caption{Descriptive statistics for years from 2003 to 2011}
\centering
    \begin{threeparttable}
    \begin{tabular}{clccccccccc}
    \toprule
    \textbf{Category} & \multicolumn{1}{l}{\textbf{Description}} & \multicolumn{9}{c}{\textbf{Mean (Standard Deviation)}} \\
    \midrule
    &  & \textbf{2003} & \textbf{2004} & \textbf{2005} & \textbf{2006} & \textbf{2007} & \textbf{2008} & \textbf{2009} & \textbf{2010} & \textbf{2011}\\
    \midrule
		Crashes & \multicolumn{1}{l}{Crash count} & 21.2 (33.8) & 17.3 (25.6) & 19.8 (27.4) & 19.4 (26.9) & 19.2 (25.8) & 15.2 (20.6) & 14.2 (19.2) & 17.4 (23.7) & 14.5 (21.6)\\
    \midrule
		Traffic & \multicolumn{1}{l} {Annual Average Daily Traffic} & 44706 (32102) & 49566 (35998) & 50820 (38000) & 50098 (36601) & 51049 (37334) & 51911 (37074) & 52504 (37918) & 50889 (36505) & 51047 (36330)\\
          & \multicolumn{1}{l}{Traffic load estimate} & 130 (12) & 173 (20) & 119 (16) & 142 (10) & 166 (14) & 167 (14) & 157 (12) & 187 (15) & 185 (15)\\
          & \multicolumn{1}{l}{Truck traffic percentage} & 11.4 (7.3) & 11.1 (6.8) & 11.1 (6.3) & 11.2 (6.3) & 10.5 (6.7) & 10.4 (6.5) & 10.7 (6.6) & 10.8 (6.4) & 10.6 (5.2)\\
          & \multicolumn{1}{l}{Imposed speed limit (miles/hours)} & 62 (6) & 59 (6) & 61 (6) & 61 (5) & 61 (5) & 61 (5) & 61 (5) & 61 (5) & 61 (5)\\
    \midrule
		Geometric & \multicolumn{1}{l}{Number of lanes (per direction)} & 3 (1) & 3 (1) & 3 (1) & 3 (1) & 3 (1) & 3 (1) & 3 (1) & 3 (1) & 3 (1) \\
          & \multicolumn{1}{l}{Total surface roadway width (ft)} & 52.5 (14.3) & 53.3 (14.9) & 54.2 (15.4) & 54.3 (15.4) & 54.5 (15.2) & 55 (15.3) & 56 (15.6) & 56.3 (15.4) & 56.2 (15.4)\\
          & \multicolumn{1}{l}{Left shoulder width (ft)} & 8.4 (2.4) & 8.2 (2.6) & 8.4 (2.7) & 8.5 (2.8) & 8.6 (2.8) & 8.7 (2.8) & 8.4 (3.3) & 8.5 (3.2) & 8.3 (3.5) \\
          & \multicolumn{1}{l}{Right shoulder width (ft)} & 8.8 (2.3) & 8.9 (2.3) & 9 (2.3) & 9 (2.3) & 9 (2.3) & 9.1 (2.3) & 9.8 (2.1) & 9.6 (2.3) & 9.5 (2.6) \\
          & \multicolumn{1}{l}{Segment length (mile)} & 0.5 (0.1) & 0.5 (0.1) & 0.5 (0.1) & 0.5 (0.1) & 0.5 (0.1) & 0.5 (0.1) & 0.5 (0.1) & 0.5 (0.1) & 0.5 (0.1)\\
    \midrule
		Pavement & \multicolumn{1}{l}{Road Condition Index} & 82 (24) & 83 (22) & 84 (20) & 84 (21) & 85 (20) & 87 (18) & 86 (19) & 88 (18) & 88 (17) \\
          & \multicolumn{1}{l}{Road Distress Index} & 88 (21) & 88 (19) & 90 (16) & 90 (17) & 92 (15) & 93 (14) & 93 (14) & 94 (13) & 94 (12)\\
          & \multicolumn{1}{l}{Road Ride Index} & 3.4 (0.6) & 3.5 (0.5) & 3.4 (0.6) & 3.5 (0.6) & 3.4 (0.6) & 3.5 (0.6) & 3.5 (0.6) & 3.5 (0.6) & 3.5 (0.6) \\
          & \multicolumn{1}{l}{Average IRI  (inch/mile)\tnote{a} } & 118 (35) & 113 (33) & 118 (36) & 114 (37) & 117 (36) & 114 (34) & 117 (36) & 113 (34) & 113 (34)\\
          & \multicolumn{1}{l}{Left wheel path IRI (inch/mile)\tnote{a}} & 117 (35) & 111 (32) & 116 (36) & 112 (36) & 117 (35) & 108 (34) & 115 (35) & 112 (33) & 107 (31)\\
          & \multicolumn{1}{l}{Right wheel path IRI (inch/mile)\tnote{a}} & 119 (36) & 116 (35) & 120 (36) & 116 (37) & 118 (38) & 121 (38) & 119 (39) & 114 (36) & 119 (40) \\
          & \multicolumn{1}{l}{Maintenance Cost (\$ scaled to hide actual budgets)} & 921 (2026) & 1133 (2191) & 834 (1286) & 887 (1834) & 1446 (9154) & 991 (1969) & 993 (1963) & 879 (1632) & 943 (2020)\\
          & \multicolumn{1}{l}{Indicator: Asphalt pavement} & 0.21 & 0.19 & 0.20 & 0.19 & 0.17 & 0.17 & 0.14 & 0.12 & 0.12 \\
          & \multicolumn{1}{l}{Indicator: CRCP pavement\tnote{b} } & 0.65 & 0.68 & 0.68 & 0.69 & 0.71 & 0.72 & 0.75 & 0.79 & 0.81\\
          & \multicolumn{1}{l}{Indicator: JCP pavement\tnote{c}} & 0.14 & 0.13 & 0.13 & 0.12 & 0.12 & 0.11 & 0.11 & 0.08 & 0.07 \\
          & \multicolumn{1}{l}{Indicator: Asphalt shoulder} & 0.61 & 0.61 & 0.60 & 0.60 & 0.60 & 0.58 & 0.58 & 0.57 & 0.57\\
    \midrule
		Location & \multicolumn{1}{l}{Indicator: Facility-interstate highway} & 0.45 & 0.45 & 0.45 & 0.45 & 0.45 & 0.45 & 0.45 & 0.45 & 0.45\\
          & \multicolumn{1}{l}{Indicator: Facility-state highway} & 0.15 & 0.15 & 0.15 & 0.15 & 0.15 & 0.15 & 0.15 & 0.15 & 0.15 \\
          & \multicolumn{1}{l}{Indicator: Facility-US highway} & 0.26 & 0.26 & 0.26 & 0.26 & 0.26 & 0.26 & 0.26 & 0.26 & 0.26\\
          & \multicolumn{1}{l}{Indicator: Rural Area} & 0.28 & 0.27 & 0.27 & 0.27 & 0.27 & 0.27 & 0.27 & 0.20 & 0.20 \\
    \bottomrule
    \end{tabular}	
		\begin{tablenotes}
            \item[a] IRI: international roughness index
            \item[b] CRCP: continuously reinforced concrete pavements
            \item[c] JCP: jointed concrete pavement
        \end{tablenotes}
        \end{threeparttable}
 \label{tab:desc_time}
\end{table}
\vspace*{\fill}
\end{landscape}

\begin{table}[ht!]
  \centering
  \footnotesize
  \caption{Posterior estimation results for dynamic spatial NB (DSNB) model}
    \resizebox{1\textwidth}{!}{\begin{tabular}{rcccc}
    \toprule
    \multicolumn{1}{l}{\textbf{Description }} & \textbf{Posterior mean} & \textbf{Posterior Std.Dev.} & \textbf{2.5\%-quantile} & \textbf{97.5\%-quantile} \\ 
    \midrule
    \multicolumn{5}{c}{Time-invariant parameters} \\
    \midrule
		\multicolumn{1}{l}{Intercept} & 0.801  &  0.077  &  0.609  &  0.894 \\
    \multicolumn{1}{l}{Indicator Variable: Asphalt pavement} &  0.271  &  0.044  &  0.192  &  0.369 \\
    \multicolumn{1}{l}{Indicator Variable: Facility-IH} & 0.712  &  0.066  &  0.601  &  0.847 \\
    \multicolumn{1}{l}{r} & 1.416  &  0.041  &  1.339  &  1.488 \\
    \midrule
		\multicolumn{5}{c}{AR (1) variance for time-varying parameters} \\
    \midrule
	\multicolumn{1}{l}{Intercept} & 0.028  &  0.023  &  0.006  &  0.084 \\
	\multicolumn{1}{l}{Annual Average Daily Traffic}  & 0.012  &  0.021  &  0.001  &  0.066 \\
	\multicolumn{1}{l}{Segment length (mile)} & 0.002  &  0.002  &  0.000  &  0.008 \\
    \multicolumn{1}{l}{Truck traffic percentage} & 0.013  &  0.016  &  0.001  &  0.055 \\
    \multicolumn{1}{l}{Indicator Variable: Rural Area} & 0.020  &  0.022  &  0.002  &  0.081 \\
    \multicolumn{1}{l}{Indicator Variable: Shoulder type - Asphalt} & 0.006  &  0.007  &  0.001  &  0.024 \\
    
    
    \multicolumn{1}{l}{Avg International Roughness Index (IRI inch/mile)} & 0.003  &  0.004  &  0.000  &  0.011 \\
    \multicolumn{1}{l}{Total shoulder width (ft)} & 0.004  &  0.006  &  0.001  &  0.019 \\
    \multicolumn{1}{l}{Imposed speed limit (miles/hour)} & 0.003  &  0.003  &  0.000  &  0.009 \\
    \midrule
    \multicolumn{4}{c}{MCMC diagnostics and sample meta data} \\
    \midrule
    \multicolumn{1}{l}{Deviance Information Criterion (DIC)} & 69113.02 \\
    \multicolumn{1}{l}{Number of MCMC iterations} & 3000 \\
    \multicolumn{1}{l}{Number of burnin iterations} & 2000 \\
    \multicolumn{1}{l}{Number of road segments} & 1158 \\
    \multicolumn{1}{l}{Number of years of data} & 9 \\
    \bottomrule
    \end{tabular}}%
  \label{tab:dynpost}%
\end{table}%

\subsection{Discussion}
\begin{figure}[ht!]
\centering
\includegraphics[width=0.95\textwidth]{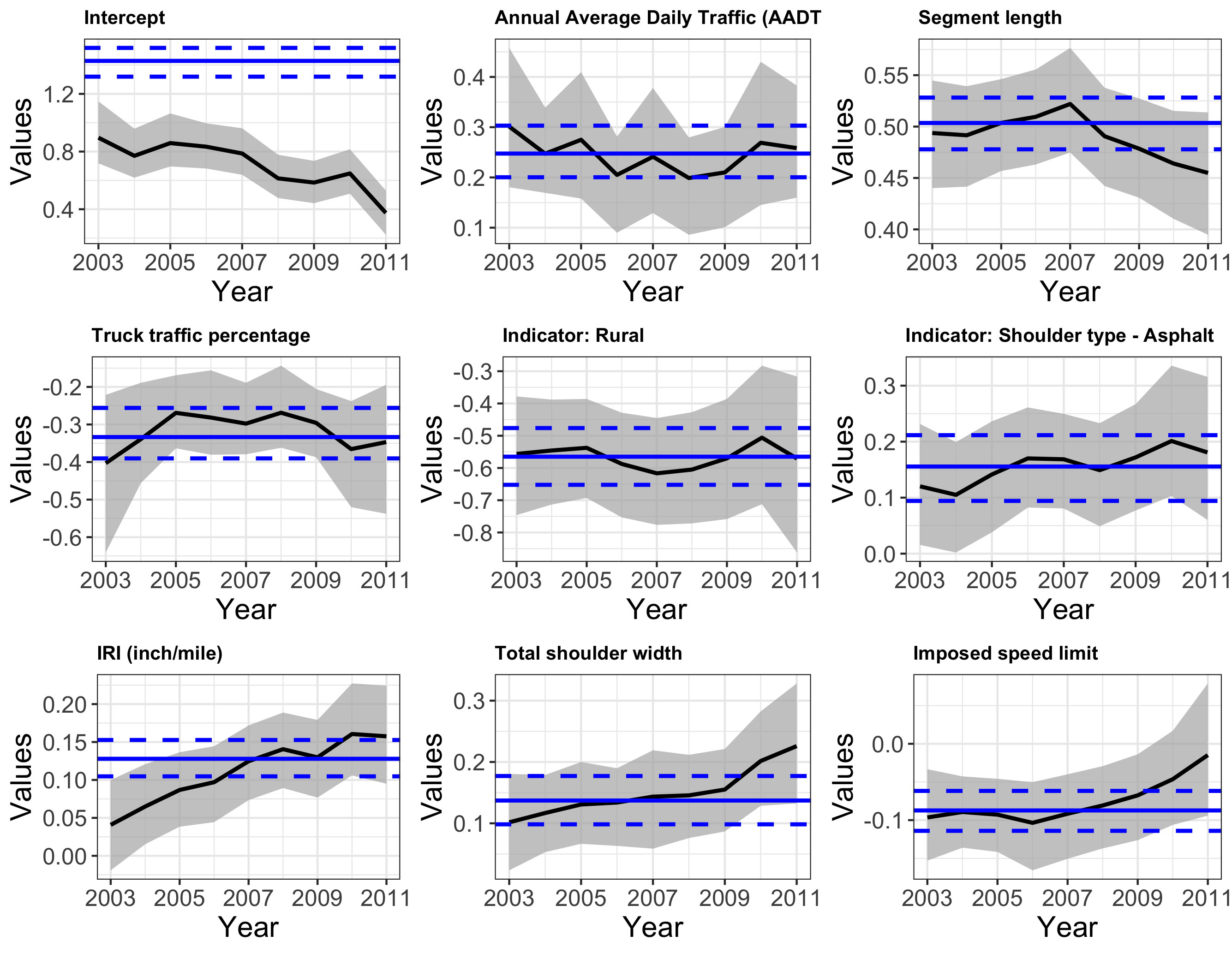}
\vspace{-0.4cm}\caption{Posterior estimates of temporal parameters from dynamic spatial NB (DSNB) specification (Black lines: posterior mean; shaded region: 95\% credible  intervals) and spatial NB (SNB) specification (Blue lines: posterior mean; dashed blue lines: 95\% credible intervals)}
\label{fig:emp_results}
\end{figure}

\begin{figure}[ht!]
\centering
\includegraphics[width=0.55\textwidth]{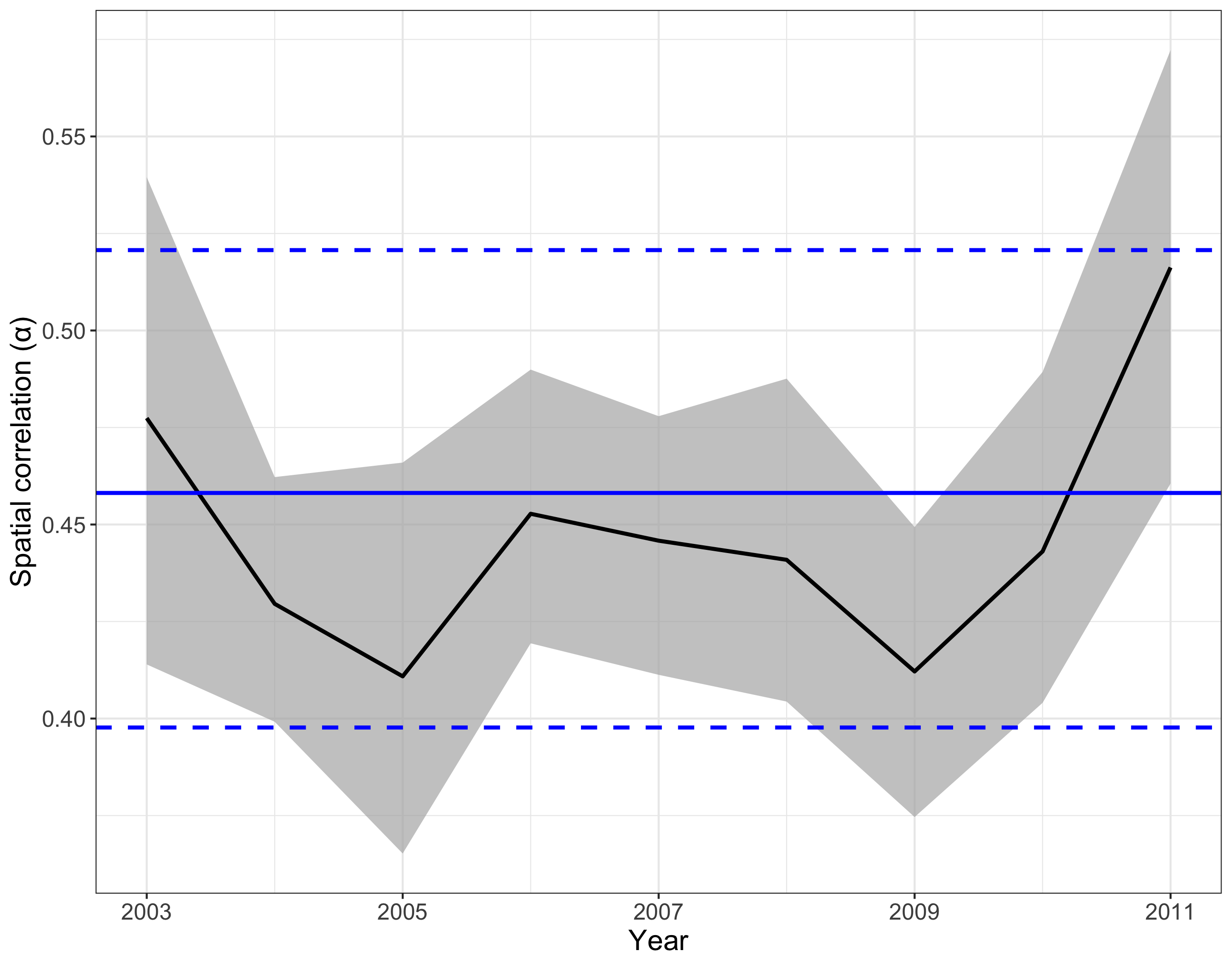}
\vspace{-0.4cm}\caption{Temporal variation in spatial correlation}
\label{fig:sp_cor}
\end{figure}

We mainly focus on the findings of the DSNB model and compare them with those of the SNB specification. Figure \ref{fig:emp_results} shows that except intercept, 95\% credible intervals of time-varying parameters in DSNB model almost fully cover those in SNB model. We observe a similar overlapping pattern for spatial correlation in Figure \ref{fig:sp_cor}. This figure also shows a little temporal variation in the magnitude of spatial correlation. The AR(1) variance estimates of time-varying parameters in Table \ref{tab:dynpost} indicate that many link function parameters in DSNB specification exhibit statistically significant temporal instability 
and trends in Figure \ref{fig:emp_results} shows that the temporal changes in many of these parameters are practically significant. We now discuss sign, magnitude, and practical implications of parameter estimates.     

The positive posterior mean estimates indicate that road segments on interstate highways with asphalt pavements are likely to have higher mean crash counts as compared to those on other facility types with concrete pavements, keeping all other covariates constant. These differences between pavement-facility types remain constant over time because corresponding indicators have time-invariant parameters in the final specification.

Among time-varying parameters, we first discuss segment length and AADT, which are generally considered as exposure variables. As expected, the longer segments with higher AADT are associated with higher mean crash counts in each year. The mean crash count per unit increase in segment length declines after 2007, and the effect of AADT on crash counts remains nearly constant (see Figure \ref{fig:emp_results}). The magnitude of the negative association of the proportion of truck traffic volume with mean crash count first decreases, then remains constant, and finally increases during the study period. Negative posterior mean estimate suggests that road segments in rural areas experience lower mean crash counts as compared to that of non-rural road segments. The difference in mean crash counts of rural and non-rural road segments remains fairly constant over the years. The pavements with asphalt shoulders experience larger mean crash count relative to the other types of shoulders throughout the study period, and the magnitude of this difference temporally increases as shown in the Figure \ref{fig:emp_results}. 
International roughness index (IRI) is positively associated with the mean crash counts, and the strength of this association consistently increases with time (see Figure \ref{fig:emp_results}). Since lower IRI values are proxy for superior road conditions, this result indicates that the safety benefits of improving the ride quality of road segments are increasing over time. 

\clearpage

\clearpage

The shoulder width also has a similar trend of its relationship with crash counts as of IRI. This result is a good illustration of the risk compensating behaviour of road users -- they may feel safer while driving on the road segments with larger shoulder widths, and therefore, they may be less careful due to perception of lower risk. Our posterior analysis also suggests that the road segments with higher speed limits are associated with lower mean crash counts during the study period, ceteris paribus, and the safety impact of a marginal increase in speed limit reduces over time. Increase in road congestion, a potential unobserved confounder, over time could be the reason behind the reduction in the influence of speed limit on crash counts. In summary, the empirical findings of the proposed DSNB model and comparison with those of the SNB model highlight the importance of specifying time-varying parameters in crash count data models. 

\section{Conclusions and Future Work} \label{sec:conc}

We propose a dynamic spatial negative binomial count data model that simultaneously allows for time-varying parameters using a dynamic linear model formulation, unobserved heterogeneity in time-invariant parameters, and time-varying spatial correlations. This specification provides an elegant way to evaluate temporal instability, a term coined by \cite{mannering2018temporal}, in model parameters while accounting for all potential sources of heterogeneity. Estimating such a flexible model using traditional Markov Chain Monte Carlo methods is challenging due to non-conjugacy of the negative binomial likelihood. To this end, we leverage recently developed P{\'o}lya-gamma data-augmentation technique to address the challenges associate with non-conjugacy and adopt the Forward Filtering and Backward Sampling algorithm to perform posterior inference on dynamic parameters. These advancements enable us to derive full conditional distributions for all the model parameters, and thus, obviate the need of Metropolis-Hasting step, resulting in a computationally-efficient and robust Gibbs sampler for the proposed model. We demonstrate the finite sample and convergence properties of the proposed Gibbs sampler in a comprehensive simulation study. 

We also demonstrate the advantages of the proposed specification in modeling crash frequency spanning across nine years (from 2003 to 2011) from a freeway road network of Houston city, USA. For this analysis, we create annually-aggregated features of road segments by fusing road management databases with crash frequency information. A majority of parameters corresponding to both time-varying and time-invariant explanatory variables exhibit statistically significant temporal instability. For example, the increasing magnitude of the positive parameter corresponding to the road roughness index suggests the temporal increase in the importance of maintaining superior ride quality to improve safety. 

While the findings from this small-scale empirical study are valuable for practitioners, our proposed framework is applicable to a higher number of features and larger networks due to the efficiency and robustness of the derived Gibbs sampler. The posterior predictive distributions of the model parameters can be constructed to predict the crash counts in a future year. However, a long crash history would be necessary to accurately predict the model parameters corresponding to a future year using a dynamic linear model. Representing unobserved heterogeneity in time-invariant parameters with a finite mixture of Gaussian distributions would enable practitioners to classify road segments in various categories in a data-driven manner \citep{buddhavarapu2016modeling} while simultaneously learning temporal evolution of other parameters, particularly for large road networks with several years of crash data. In our future work, we plan to explore recent developments in approximate Bayesian inference \citep{bansal2020fast, luts2015variational} to make the estimation even faster and scalable, and eventually facilitate periodical data-driven safety strategy development on large-scale networks for transportation agencies. 

\newpage
\bibliographystyle{elsarticle-harv}
\bibliography{bibliography}

\clearpage

\begin{appendices}

\section{P{\'o}lya-Gamma data augmentation} \label{app:polya}

A random variable $\omega$ has a PG(b,c) distribution (P{\'o}lya-Gamma distribution with parameters b and c) if
\begin{equation}\label{eq:PG random variable}
\omega \stackrel{D}{=} \frac{1}{2\pi^2}\sum\limits_{k=1}^{\infty}\frac{g_k}{(k-1/2)^2+c^2/(4\pi^2)}
\end{equation}
where, $g_1,g_2,...g_k,...$ are independent and identically distributed (i.i.d) random variables with Gamma$(b,1)$. The P{\'o}lya-Gamma distributed random variables can be generated from an infinite sum of weighted i.i.d Gamma distributed random variables. We refer the interested readers to \citet{polson_bayesian_2013} for details on the sampling methods and efficiency of drawing P{\'o}lya-Gamma random variables. 

We now discuss how a NB likelihood can be converted to a Gaussian likelihood using P{\'o}lya-gamma data augmentation. Consider a simplistic representation of the cross-sectional NB regression model:

\begin{equation} \label{NBLH:cross}
\begin{aligned}
y_{i} &\sim \text{NB}(r, p_{i}); \quad i \in \{1,2,...n\}  \\
p_{i} &= \frac{\exp(\psi_{i})}{1+\exp(\psi_{i})}; \quad \psi_{i}= X_{i}\gamma + \phi_{i} 
\end{aligned}
\end{equation}

The NB likelihood parametrized by log-odds can be written as, 
\begin{equation} 
\begin{split}
P(y_i|\psi_i,r) 
& = \frac{\Gamma{(y_i+r)}}{\Gamma{(r)}y_i!} \frac{\exp(\psi_i)^{y_i}}{(1+\exp(\psi_i))^{r+y_i}}  
\end{split}
\end{equation}

After applying the main result of \citet{polson_bayesian_2013}, the equation can be written as
\begin{equation} \label{eq:NB_to_PG}
\begin{split}
P(y_i|\psi_i,r) 
& = \frac{\Gamma{(y_i+r)}}{\Gamma{(r)}y_i!} 2^{-(r+y_i)}\exp\left(\frac{(y_i-r)\psi_i}{2}\right) \int_0^\infty \exp\left(\frac{-\omega_i \psi_i^2}{2}\right) p(\omega_i) d\omega_i \\
& = \frac{\Gamma{(y_i+r)}}{\Gamma{(r)}y_i!} 2^{-(r+y_i)}\exp\left(\frac{(y_i-r)\psi_i}{2}\right) \mathbb{E}_{\omega_i}\left[\exp\left(\frac{-\omega_i \psi_i^2}{2}\right)\right] \\
\end{split}
\end{equation}

where $ \omega_i \sim PG(y_i+r,0)$. After algebraic rearrangement of terms, the NB likelihood (a function of $\gamma$) becomes Gaussian likelihood conditional on the P{\'o}lya-Gamma random variable $\omega_i$, r and $\psi_i$ as shown below. 
\[
\begin{split}
P(y|\psi,r,\omega) 
& = \prod\limits_{i=1}^{n} P(y_i|\psi_i,r,\omega_i) \\
& = \prod\limits_{i=1}^{n} \frac{\Gamma{(y_i+r)}}{\Gamma{(r)}y_i!} 2^{-(r+y_i)} \exp\left(\frac{(y_i-r)\psi_i}{2}\right) \exp\left(\frac{-\omega_i \psi_i^2}{2}\right)\\
& \propto \prod\limits_{i=1}^{n} 
\exp\left(- \frac{\omega_{i}}{2}\left[\psi_{i} - \frac{y_{i}-r}{2\omega_{i}} \right]^2\right)\\
& \propto \exp\left(-\frac{(z-\psi)^{\prime} \Omega (z-\psi) }{2}\right)
\end{split}
\]
where, $z_i = \frac{y_i-r}{2\omega_i}$ for $i \in \{1,2,....n\}$  and $\psi_i$ contains the regression coefficients. 
\[
\begin{split}
    z_{i} &=  \psi_{i} + \nu_{i}; \quad \nu_i \sim \text{N}(0,\omega_{i}^{-1})\\
    z &=  \psi + \nu; \quad \nu \sim \text{Normal}(0,\Omega)
\end{split}
\]

where $\Omega = \text{diag}(\omega_{i}^{-1})$. In the Gibbs sampler, we pretend that $z_i$ is observed instead of $y_i$ and thus, a Gaussian likelihood form of crash counts is obtained in terms of $z$. We also consider a panel data setting, where $\Omega_t$ and  $\Omega_i$ are used to represent time-specific and individual-specific variance-covariance matrices of $n \times n$ and $T \times T$, respectively.

\section{Gibbs Sampler} \label{app:gibbs}
This section provides a Gibbs sampling scheme to iteratively draw from full conditional posterior distributions of the parameters of the proposed dynamic spatial NB model with dynamic parameters and random heterogeneity.

\subsection*{\textbf{B.1 Posterior sampling of the dispersion parameter} ($r$)} 

To derive posterior distribution of the dispersion parameter $r$, we express NB random variables as sums of Logarithmic random variables under compound Poisson distribution \citep{quenouille_relation_1949}:
\begin{equation}\label{NB compound Poisson} 
y_{it}  = \sum\limits_{k=1}^{L_{it}}\varsigma_{kt}, \indent \varsigma_{kt} \overset{iid}{\sim} \text{Logarithmic}(p_{it}), \indent L_{it} \sim \text{Poisson}(-r\ln(1-p_{it})), \indent \forall i \in \{1,...,n\}, \; t \in \{1,...,T\}  
\end{equation}
\begin{equation} \label{eq:r}
\begin{split}
P(r \lvert L, p, \text{\textendash}) 
& \propto \prod\limits_{t=1}^{T}\prod\limits_{i=1}^{n} P(L_{it} | r) P(r) \\
r \lvert  L, p, \text{\textendash} & \sim \text{Gamma}\left(r_0+\sum\limits_{t=1}^{T}\sum\limits_{i=1}^{N}L_{it}, h-\sum\limits_{t=1}^{T}\sum\limits_{i=1}^{N}\ln(1-p_{it})\right) \\
\end{split}
\end{equation}

The conditional posterior of $L_{it}$ is obtained using the procedure described in \cite{zhou_lognormal_2012}. Here, we provide the analytical closed form expression
for the posterior of discrete random variable  $L_{it}$ below:
\begin{equation} \label{eq:L}
P(L_{it} = j \lvert r,y) = R(y_{it},j), \indent j \in \{0, 1,...y_{it}\}
\end{equation}
where, 
\[ 
R(l,m) =  
\begin{cases} 
		1 & l = 0; m = 0 \\
		\\
		\frac{F(l,m)r^m}{\sum\limits_{j=1}^{l}F(l,j)r^{j}} & l \neq 0 ; m \neq 0 \\      
 \end{cases}
\]
\[ 
F(m,j) =  
\begin{cases} 
		1 & m = 1 \; \& \; j = 1 \\
		0 & m < j \\      
		\frac{(m-1)}{m}F(m-1,j)+\frac{1}{m}F(m-1,j-1) & 1\leq j\leq m
 \end{cases}
\]
The hyper parameter $h$ is learned by creating the full conditional posterior:
\begin{equation} \label{eq:H}
\begin{split}
    P(h \lvert  r, \text{\textendash}) & \propto P(r\lvert r_{0},h) P(h \lvert e_{0},f_{0}) \\
    h \lvert  r, \text{\textendash} & \sim  \text{Gamma}(r_{0} + he_{0},  r + hf_{0}) 
\end{split}    
\end{equation}

\subsection*{\textbf{B.2 Posterior sampling of $\{\beta_{1:n}, \mu_{1:C}, \gamma, \omega_{\{1:n,1:T\}}\}$}}
Using the P{\'o}lya-Gamma data augmentation, the crash counts $y_{it}$ are transformed to $z_{it} = \frac{y_{it}-r}{2\omega_{it}}$ as described in Appendix \ref{app:polya}. Pretending that $z_{it}$ is observed, instead of $y_{it}$, we obtain a Gaussian likelihood form of crash counts in terms of $z_{it}$. We sample the time-invariant random parameters $\beta_{1:n}$, their component-specific means $\mu_{1:C}$, fixed parameters $\gamma$, and auxiliary variable $\omega$ in blocks to accelerate the convergence by improving the mixing of Markov chains. Considering the joint distribution $p(\beta_{1:N},\gamma, \mu_{1:C}, \omega \lvert \text{\textendash})$ given in equation \ref{eq:joint}, we perform this blocked sampling in the following steps: 

\begin{equation} \label{eq:joint}
    p(\beta_{1:N},\gamma, \mu_{1:C}, \omega \lvert \text{\textendash}) \propto \prod\limits_{i=1}^{n} p(\beta_{i} \lvert \gamma, \mu_{1:C}, \omega,  \text{\textendash}) p(\gamma, \mu_{1:C} \lvert \omega, \text{\textendash}) p(\omega \lvert \text{\textendash})
\end{equation}

\begin{itemize}
    \item I: $p(\beta_{i} \lvert \gamma, \mu_{1:C}, \omega,  \text{\textendash}) \quad \forall i \in \{1,\dots,n\}$  
    \item II: $p(\gamma, \mu_{1:C} \lvert \omega, \text{\textendash})$
    \item III: $p(\omega \lvert \text{\textendash})$
\end{itemize}

\subsubsection*{I: Sampling from $p(\beta_{i} \lvert \gamma, \mu_{1:C}, \omega, \text{\textendash})$}

\begin{equation}\label{eq:beta_i}
\begin{split}
P(\beta_i \lvert \gamma, \mu_{1:C}, \omega,  \text{\textendash}) 
& \propto \prod\limits_{t=1}^{T} P(z_{it}|\psi_{it},\omega,r) P(\beta_i \lvert \varrho_{i})\\
& \propto \exp\left({-\frac{1}{2}(z_i-\psi_i)^{\prime}\Omega_{i}^{-1}((z_i-\psi_i)}\right)
 \exp\left(\prod\limits_{c=1}^{C}{\left({-\frac{1}{2}(\beta_i-\mu_c)^{\prime}\Sigma_c^{-1}(\beta_i-\mu_c)}\right)}^{I(\varrho_i=c)}\right)  \\
& \sim \text{Normal}(m_{\beta_i}, V_{\beta_i})  \\
\text{where} \quad \quad \quad V_{\beta_i} & = \left[{X^{R}_{i}}^{\prime}\Omega_i^{-1} {X^{R}_{i}} + \prod\limits_{c=1}^{C}\left(\Sigma_c^{-1}\right)^{I(\varrho_i=c)}\right]^{-1} \\
m_{\beta_i} & = V_{\beta_i} \left[{X^{R}_{i}}^{\prime}\Omega_i^{-1}(z_i-{X^{F}_{i}}\gamma-{X^{D}_{i}}\theta_{t} -\phi_i) + \prod\limits_{c=1}^{C}{(\Sigma_c^{-1}\mu_c)}^{I(\varrho_i=c)}\right] \end{split}
\end{equation}

where $X^{F}_{i}$,  $X^{R}_{i}$, and $X^{D}_{i}$ are covariates of $T \times g$, $T \times h$ and $T \times q$ dimension, respectively; $\Omega_{i}$ is a $T \times T$ diagonal matrix with diagonal elements $\omega_{it}^{-1}$ $\forall t \in \{1,2,....T\}$; $\varrho$ is a $n \times 1$ vector of latent class indicators, $\varrho_i = c$, if $i^{th}$ unit belongs to $c^{th}$ latent class; $I(\varrho_i=c)$ is atomic latent class indicator which equals one if $i^{th}$ unit belongs to $c^{th}$ latent class, else switches to zero; $z_i$, $\psi_i$, and $\phi_i$ are $T\times 1$ vectors.

\subsubsection*{II: Sampling from $p(\gamma, \mu_{1:C} \lvert \omega, \text{\textendash})$}
To construct the conditional posterior $p(\gamma, \mu_{1:C} \lvert \omega, \text{\textendash})$, we marginalize the random parameters as follows.

\begin{equation*}
\begin{split}
z_i & = {X^{F}_{i}}\gamma + {X^{R}_{i}}\beta_i + X^{D}_{i}\theta_{t} + \phi_i + \nu_{i}; \quad \nu_{i} \sim \text{Normal}(0,\Omega_i) \\
 & = {X^{F}_{i}}\gamma + \sum\limits_{c=1}^{C}I(\varrho_i=c) {X^{R}_{i}} \mu_c +  X^{D}_{i}\theta_{t}  + \phi_i + \nu_{i}^{*}; \quad \nu_{i}^{*} \sim \text{Normal}(0,\Upsilon_i), \quad \Upsilon_i = \Omega_i + \sum\limits_{c=1}^{C}I(\varrho_i=c) {X^{R}_{i}} \Sigma_c {X^{R}_{i}}^{\prime} 
\end{split}
\end{equation*}

The joint distribution can be written as: 
\[
p(\gamma, \mu_{1:C} \lvert \varrho, \omega, \text{\textendash}) \propto \prod\limits_{c=1}^{C} p(\mu_c \lvert  \gamma,\omega, \varrho, \text{\textendash}) p(\gamma \lvert \omega, \text{\textendash})
\]

\begin{itemize}
\item II-1: $\mu_c \lvert  \gamma,\omega, \text{\textendash} \sim \text{Normal}(m_{\mu_c}, V_{\mu_c})$, where 

\begin{equation} \label{eq:mu}
\begin{split}
     V_{\mu_c} & =  \left(\sum\limits_{i=1}^{n} {X^{R}_{i}}^{\prime} \Upsilon_{i}^{-1} X^{R}_{i} I(\varrho_i=c) + B_0^{-1}\right)^{-1}\\
    m_{\mu_c}(\gamma) & = V_{\mu_c} \left(\sum\limits_{i=1}^{n} {X^{R}_{i}}^{\prime} \Upsilon_{i}^{-1} I(\varrho_i=c) (z_i-\phi_i-{X^{F}_{i}} \gamma - X^{D}_{i} \theta_{t}) + B_0^{-1}b_0\right)
\end{split}    
\end{equation}

\item II-2: $\gamma \lvert \omega, \text{\textendash} \sim \text{Normal}(m_{\gamma}, V_{\gamma})$,  where 

\begin{equation} \label{eq:gamma}
\begin{split}
    V_{\gamma} & =   \left(\sum\limits_{i=1}^{n} {X^F_i}^{\prime} \Upsilon_i^{-1}(\Upsilon_i-X_i^R V_{\mu_{\varrho_i}}{X^R_i}^{\prime}) \Upsilon_i^{-1} X^F_i + S_0^{-1}\right)^{-1}\\
    m_{\gamma} & = V_{\gamma} \left(\sum\limits_{i=1}^{n} {X^F_i}^{\prime} \Upsilon_i^{-1} (z_i-\phi_i - X_i^R m_{\mu_{\varrho_i}}(0) - X_i^D \theta_{t}) + S_0^{-1}s_0\right)
\end{split}    
\end{equation}
\end{itemize}

\subsubsection*{III: Sample from $p(\omega \lvert \text{\textendash})$}
The full conditional distribution $\omega \lvert \text{\textendash}$ turns out to be a distribution in the P{\'o}lya-Gamma class \citep[see][for details]{polson_bayesian_2013}. 
\begin{equation} \label{eq:omega}  \omega_{it} \lvert \text{\textendash} \sim \text{PG}(y_{it}+r, \psi_{it})
\end{equation}

The details of the P{\'o}lya-Gamma distribution are provided in Appendix \ref{app:polya}. 

\subsection*{\textbf{B.3 Posterior sampling from component-specific variance ($\Sigma_{1:C}$})} 
\begin{equation}\label{eq:sigma_c}
\begin{split}
P(\Sigma_c \lvert \text{\textendash})
& \propto \prod\limits_{i=1}^{n} P(\beta_{i} \lvert \mu_c,\Sigma_c,\varrho_i) P(\Sigma_c)\\
\Sigma_c \lvert \text{\textendash} & \sim \text{Wish}(\nu_{\Sigma_c}, V_{\Sigma_c}) \\
\text{where} \quad \quad \quad \nu_{\Sigma_c} & = \nu_0 + \sum\limits_{i=1}^{n} I(\varrho_i=c) \\
V_{\Sigma_c} & = \left(V_0^{-1} + \sum\limits_{i=1}^{n} I(\varrho_i=c) (\beta_i-\mu_c)(\beta_i-\mu_c)^{\prime}\right)^{-1}
\end{split}
\end{equation}

\subsection*{\textbf{B.4 Posterior sampling from categorical variable ($\varrho_{1:n}$)}} 

\begin{equation}\label{eq:latent}
P(\varrho_i = c \lvert \text{\textendash})
=  \frac{\Phi(z_{i};m_{\varrho_{ic}},V_{\varrho_{ic}}) \eta_c}{\sum\limits_{c^{\prime}=1}^{C} \Phi(z_{i};m_{\varrho_{ic^{\prime}}},V_{\varrho_{ic^{\prime}}}) \eta_c^{\prime}}
\end{equation}

Where, $m_{\varrho_{ic}} = {X^{F}_{i}}\gamma + {X^{R}_{i}}\mu_c + {X^{D}_{i}}\theta_{t} + \phi_i$,  $V_{\varrho_{ic}} = X^{R}_{i} \Sigma_c {X^{R}_{i}}^{\prime} + \Omega_i$, $\Phi(.)$ is a multivariate Gaussian density function, and $\eta=[\eta_1, \eta_2,.....,\eta_C]$ is the mixture weight vector. We sample $\eta$ using the following full conditional distribution. 

\begin{equation} \label{eq:eta}
\begin{split}
P(\eta \lvert \varrho)
& \propto P(\varrho \lvert \eta) P(\eta)\\
\eta \lvert \varrho & \sim  \text{Dirichlet}(\alpha_0+g_1, \alpha_0+g_2,.....,\alpha_0+g_C)
\end{split}
\end{equation}
where, $g_c = \sum\limits_{i=1}^{n} (\varrho_i=c)$.

\subsection*{\textbf{B.5 Posterior sampling of state vector ($\theta_{t}$)}}

The evolution equations of the system are written as follows using the transformed data $\zeta_t$. 
\begin{equation*} 
\begin{split}
\zeta_t &= F_t \theta_t + \nu_t; \quad \nu_t \sim \text{Normal}(0,\Omega_t) \\
\theta_t &= G_t \theta_{t-1} + u_t;  \quad  u_t \sim \text{Normal}(0,W_t)
\end{split}
\end{equation*}

FFBS algorithm is performed in two steps: Forward filtering and backward smoothing. The following recursions are performed in each MCMC iteration within the Gibbs sampler.

\subsubsection*{\textbf{B.5.1 Forward Filtering}} 
The following recursions are performed in each MCMC iteration within the Gibbs sampler. We start the recursions by drawing a state vector $\theta_0$ from a non-informative distribution, which is the state prior distribution at $t=1$ and the state posterior at time $t=0$. 

\noindent \\ Initialisation: 
\[
\theta_0 \sim \text{Normal}(m_0,C_0)
\]

\noindent \emph{\textbf{from t = 1 to T}}:\\

\noindent Posterior at t-1: 
\[
\theta_{t-1} \lvert D_{t-1} \sim \text{Normal}(m_{t-1},C_{t-1})
\]

\noindent Prior at t: 
\begin{equation*} 
\begin{split}
\theta_{t}\lvert D_{t-1} & \sim \text{Normal}(a_{t},R_{t}) \\
a_t &= G_t m_{t-1} \\
R_t &= G_t C_{t-1} G_t^{\prime} + W_t
\end{split}
\end{equation*}

\noindent Predictive at t: 

\begin{equation*} 
\begin{split}
\zeta_t \lvert D_{t-1} & \sim \text{Normal}(f_t, Q_t) \\
f_t &= F_t a_t \\
Q_t &= F_t R_t F_t^{\prime} + \Omega_t
\end{split}
\end{equation*}

\noindent Posterior at t: 
\begin{equation*} 
\begin{split}
\theta_{t} \lvert D_{t} & \sim \text{Normal}(m_{t},C_{t}) \\
m_t &= a_t + R_t F_t^{\prime} Q_t ^{-1}(\zeta_t-f_t)  \\
C_t &= R_t - R_t F_t^{\prime} Q_t^{-1} F_t R_t
\end{split}
\end{equation*}

where $D_{t}$ is the information provided by the first $t$ observations. The above computations involve inversion of $Q_t$ of size $n \times n$, which becomes computationally expensive with the number of road segments. However, we use the following established result from matrix algebra to circumvent this challenge:

\[
Q_t^{-1} = (F_t R_t F_t^{\prime} + \Omega_t)^{-1} = \Omega_t^{-1}-\Omega_t^{-1} F R (I_q + F_t^{\prime} \Omega_t^{-1} F R)^{-1} F^{\prime} \Omega_{t}^{-1}
\]

We continue the recursions until time $T$ and then draw the state vector at time $T$ using $\theta_T \lvert D_T \sim \text{Normal}(m_T, C_T)$. 

\subsubsection*{\textbf{B.5.2 Backward Smoothing }} 

Subsequently, we recursively draw the remaining states $\theta_{1:T-1}$ by backward smoothing using the following equations and then a single draw of the complete state vector ${\theta_1,..\theta_T}$ is available at each MCMC iteration:

\begin{equation*} 
\begin{split}
\theta_{t} \lvert \theta_{t+1},D_{t} & \sim \text{Normal}(h_{t},B_{t}) \\
h_t & = m_t + C_t G_{t+1}^{\prime} R_{t+1}^{-1} (h_{t+1} - a_{t+1})\\
B_t & = C_t - C_t G_{t+1}^{\prime} R_{t+1}^{-1} (R_{t+1} - B_{t+1}) R_{t+1}^{-1} G_{t+1}^{\prime} C_t
\end{split}
\end{equation*}

\subsection*{\textbf{B.6 Posterior sampling of $\sigma_{\theta k}^2$ }}
For each $k^{th}$ diagonal element $\sigma_{\theta k}^2$ of $W_{t}$, where $k \in {1,2,\ldots q}$, take a draw in each MCMC iteration from: 
\begin{equation}
\frac{1}{\sigma_{\theta k}^2} \lvert \text{\textendash}  \sim \text{Gamma}\left( a_{\sigma} +\frac{1}{2} (T-1), \frac{1}{b_{\sigma}+ \frac{1}{2}\sum\limits_{t=2}^{T} {(\theta_{t}-G_t[k,k] \theta_{t-1})}^2}\right)
\end{equation}
where, $G_t[k,k]$ is the $k^{th}$ diagonal element of $G_t$. 

\subsection*{\textbf{B.7 Posterior sampling of spatial random effects ($\phi_{i}^{t}$)}}

\begin{equation}
\begin{split}
p(\phi_i^{t} \lvert \phi_{-i}^{t}, \text{\textendash})
& \propto P(z_{it} \lvert \phi_{-i}^{t},\text{\textendash}) P(\phi_i^{t} \lvert \phi_{-i}^{t}) \\
& \propto \exp\left({\frac{-\omega_{it}}{2}\left[z_{it}-(X^{F}_{it}\gamma + X^{R}_{it}\beta_{i} + X^{D}_{it}\theta_{t} + \phi_{i}^{t})\right]^2}\right) \exp\left(-\frac{w_{i+}}{2\tau_t^2} \left[ \phi_{i}^{t} - \sum\limits_{j} \frac{w_{ij}}{w_{i+}} \phi_{j}^{t} \right]^{2}\right)  \\
\phi_i^{t} \lvert \phi_{-i}^{(t)}, \text{\textendash} & \sim \text{Normal}\left( V_{\phi}^{t} \left([z_{it}-(X^{F}_{it}\gamma + X^{R}_{it}\beta_{i} + X^{D}_{it}\theta_{t})]\omega_{it}+\left(\sum\limits_{j} w_{ij}\phi_j^{t}\right)\frac{1}{\tau_t^2}\right), V_{\phi}^{t} \right) \\
\text{where} \quad \quad \quad V_{\phi}^{t} & = \left(\omega_{it}+\frac{w_{i+}}{\tau_t^2}\right)^{-1} 
\end{split}
\end{equation}

We perform mean centering to accommodate the identification issue: $\phi^{t} = \phi^{t} - \bar{\phi^{t}}$ , where $\bar{\phi^{t}}$ is mean of the vector $\phi^{t}$. \\

\subsection*{\textbf{B.8 Posterior sampling of spatial parameter ($\tau_{t}$)}}

\begin{equation}\label{eq:tau_c_t}
\tau_t^{-2} \lvert \text{\textendash}  \sim \text{Gamma}\left(c_0+\frac{n}{2}, d_0+\sum\limits_{i=1}^{n}\frac{w_{i+}}{2}\left[ \phi_{i}^{t} - \sum\limits_{j} \frac{w_{ij}}{w_{i+}} \phi_{j}^{t} \right]^{2}\right)
\end{equation}

\newpage
\thispagestyle{empty}
\newgeometry{left=2cm,right=2cm, top=2cm, bottom = 2cm}

\begin{algorithm}[!ht]
\SetAlgoLined
\SetKwFor{For}{for}{sample}{end for}%
\textbf{Initialization:}\\
Initialize parameters: $\{ r,h, \gamma,  \beta_{i}, \mu_{c}, \Sigma_{c}, \eta_{c}, \omega_{it}, \theta_{t},  \sigma_{\theta k}^2, \phi_{i}^{t}, \tau_{t} \}, \quad \forall i \in \{1,\dots,n\}, \forall t \in \{1,\dots,T\},\forall k \in \{1,\dots,q\}, \forall c \in \{1,\dots,C\} $  \;
Set hyper-parameters: $\{r_0, e_{0}, f_{0}, s_0, S_0, b_0, B_0, \nu_0, V_0, \alpha_0, m_0, C_0, c_0, d_0 ,  a_{\sigma}, b_{\sigma}\}$    \;
\For{1 \KwTo max-iteration}{
i) $ r \lvert \text{\textendash} \sim  \text{Gamma}\left(r_0+\sum\limits_{t=1}^{T}\sum\limits_{i=1}^{N}L_{it}, h-\sum\limits_{t=1}^{T}\sum\limits_{i=1}^{N}\ln(1-p_{it})\right)\quad$ (see equation \ref{eq:L} for the details of $L_{it}$)\;
ii) $h \lvert \text{\textendash} \sim  \text{Gamma}(r_{0} + he_{0},  r + hf_{0})$ \;
iii) $\{\beta_{1:n}, \mu_{1:C}, \gamma, \omega_{\{1:n,1:T\}} \}$:
       \begin{itemize}
           \item  $\{\beta_{i} \lvert \text{\textendash} \}_{i=1}^{n}  \sim \text{Normal}(V_{\beta_i} \left[{X^{R}_{i}}^{\prime}\Omega_i^{-1}(z_i-{X^{F}_{i}}\gamma-{X^{D}_{i}}\theta_{t} -\phi_i) + \prod\limits_{c=1}^{C}{(\Sigma_c^{-1}\mu_c)}^{I(\varrho_i=c)}\right], V_{\beta_i})$, where \\
           $V_{\beta_i} = \left[{X^{R}_{i}}^{\prime}\Omega_i^{-1} {X^{R}_{i}} + \prod\limits_{c=1}^{C}\left(\Sigma_c^{-1}\right)^{I(\varrho_i=c)}\right]^{-1}$ \; 
           \item $\{\mu_c  \lvert \text{\textendash}\}_{c=1}^{C} \sim \text{Normal}(V_{\mu_c} \left(\sum\limits_{i=1}^{n} {X^{R}_{i}}^{\prime} \Upsilon_{i}^{-1} I(\varrho_i=c) (z_i-\phi_i-{X^{F}_{i}} \gamma - X^{D}_{i} \theta_{t}) + B_0^{-1}b_0\right), V_{\mu_c})$, where \\
           $V_{\mu_c} =  \left(\sum\limits_{i=1}^{n} {X^{R}_{i}}^{\prime} \Upsilon_{i}^{-1} X^{R}_{i} I(\varrho_i=c) + B_0^{-1}\right)^{-1}$ \;
           \item $\gamma \lvert \text{\textendash} \sim \text{Normal}(V_{\gamma} \left(\sum\limits_{i=1}^{n} {X^F_i}^{\prime} \Upsilon_i^{-1} (z_i-\phi_i - X_i^R m_{\mu_{\varrho_i}}(0) - X_i^D \theta_{t}) + S_0^{-1}s_0\right), V_{\gamma})$, where \\
           $V_{\gamma} =   \left(\sum\limits_{i=1}^{n} {X^F_i}^{\prime} \Upsilon_i^{-1}(\Upsilon_i-X_i^R V_{\mu_{\varrho_i}}{X^R_i}^{\prime}) \Upsilon_i^{-1} X^F_i + S_0^{-1}\right)^{-1}$\;
           \item $\left\{\omega_{it} \lvert \text{\textendash}\right\}_{\forall i,t} \sim \text{PG}(y_{it}+r, \psi_{it})$ \;
       \end{itemize}
iv) $\left\{\Sigma_{c} \lvert \text{\textendash}\right\}_{c=1}^{C} \sim \text{Wish}\left( \nu_0 + \sum\limits_{i=1}^{n} I(\varrho_i=c), \left(V_0^{-1} + \sum\limits_{i=1}^{n} I(\varrho_i=c) (\beta_i-\mu_c)(\beta_i-\mu_c)^{\prime}\right)^{-1} \right)$ \; 
v) Class indicator $\{\varrho_{i}\}_{i=1}^{n}$ using $P(\varrho_i = c \lvert \text{\textendash})
=  \frac{\Phi(z_{i};m_{\varrho_{ic}},V_{\varrho_{ic}}) \eta_c}{\sum\limits_{c^{\prime}=1}^{C} \Phi(z_{i};m_{\varrho_{ic^{\prime}}},V_{\varrho_{ic^{\prime}}}) \eta_c^{\prime}}$, where \\
\quad \quad $m_{\varrho_{ic}} = {X^{F}_{i}}\gamma + {X^{R}_{i}}\mu_c + {X^{D}_{i}}\theta_{t} + \phi_i$ and  $V_{\varrho_{ic}} = X^{R}_{i} \Sigma_c {X^{R}_{i}}^{\prime} + \Omega_i$ \;
vi) $\eta \lvert \varrho \sim  \text{Dirichlet}(\alpha_0+g_1, \alpha_0+g_2,.....,\alpha_0+g_C)$ where, $g_c = \sum\limits_{i=1}^{n} (\varrho_i=c)$ \;
vii) $\{\theta_{t}\}_{t=1}^{T}$ using forward filtering and backward smoothing (appendix B.5)\; 
\vspace{0.3cm}
Forward filtering: \\
 \For{$t$ in $1:T$}{
    Posterior at $t-1$: $\theta_{t-1} \lvert  D_{t-1} \sim \text{Normal}(m_{t-1},C_{t-1})$ \;
    Prior at $t$: $\theta_{t} \lvert  D_{t-1} \sim \text{Normal}(G_t m_{t-1},G_t C_{t-1} G_t^{\prime} + W_t)$ \;
    Predictive at $t$: $\zeta_t \lvert  D_{t-1} \sim \text{Normal}(F_t a_t, F_t R_t F_t^{\prime} + \Omega_t)$ \;
    Posterior at $t$: $\theta_{t} \lvert D_{t}  \sim \text{Normal}(a_t + R_t F_t^{\prime} Q_t ^{-1}(\zeta_t-f_t), R_t - R_t F_t^{\prime} Q_t^{-1} F_t R_t)$ ;
    \\}
\vspace{0.3cm}
Backward smoothing: \\
\For{$t$ in $(T-1):1$}{
$\theta_{t} \lvert \theta_{t+1},D_{t} \sim \text{Normal}( m_t + C_t G_{t+1}^{\prime} R_{t+1}^{-1} (h_{t+1} - a_{t+1}), C_t - C_t G_{t+1}^{\prime} R_{t+1}^{-1} (R_{t+1} - B_{t+1}) R_{t+1}^{-1} G_{t+1}^{\prime} C_t);$ \\ 
}

viii) $\left\{\frac{1}{\sigma_{\theta k}^2}\right\}_{k=1}^{q} \lvert \text{\textendash}  \sim \text{Gamma} \left( a_{\sigma} +\frac{1}{2} (T-1), \frac{1}{b_{\sigma}+ \frac{1}{2}\sum\limits_{t=2}^{T} {(\theta_{t}-G_t[k,k] \theta_{t-1})}^2}\right)$\;
ix) $\left\{\phi_i^{t} \lvert \text{\textendash}\right\}_{\{\forall i,t\}} \sim \text{Normal}\left( \left(\omega_{it}+\frac{w_{i+}}{\tau_t^2}\right)^{-1} \left([z_{it}-(X^{F}_{it}\gamma + X^{R}_{it}\beta_{i} + X^{D}_{it}\theta_{t})]\omega_{it}+\left(\sum\limits_{j} w_{ij}\phi_j^{t}\right)\frac{1}{\tau_t^2}\right), \left(\omega_{it}+\frac{w_{i+}}{\tau_t^2}\right)^{-1} \right)$ \; 
x) $\left\{\tau_t^{-2} \lvert \text{\textendash} \right\}_{t=1}^{T} \sim \text{Gamma}\left(c_0+\frac{n}{2}, d_0+\sum\limits_{i=1}^{n}\frac{w_{i+}}{2}\left[ \phi_{i}^{t} - \sum\limits_{j} \frac{w_{ij}}{w_{i+}} \phi_{j}^{t} \right]^{2}\right)$ \;
xi) Compute spatial correlation: $\{\alpha_{t}\}_{t=1}^{T} = \frac{\sigma_{\phi^{t}}}{\sigma_{\phi^{t}} + \sigma_{\epsilon^{t}}}$.
}
 \caption{Gibbs sampler for the spatial negative binomial model with time-varying and random parameters.}
 \label{algo:2}
 \end{algorithm}
\newpage
\restoregeometry

\end{appendices}

\end{document}